\documentclass[twocolumn]{aastex631}

\usepackage{bm}
\usepackage{colortbl}

\newcommand{\nfibers}{252 million}

\newcommand{\sqam}{arcmin$^{2}$}
\newcommand{\numobj}{493,012}
\newcommand{\numstars}{259,396}
\newcommand{\numgal}{74,196}
\newcommand{\numqso}{4,087}
\newcommand{\numunknown}{154,543}
\newcommand{\numbstar}{2,093}
\newcommand{\numastar}{12,040}
\newcommand{\numfstar}{24,152}
\newcommand{\numgstar}{66,552}
\newcommand{\numkstar}{133,457}
\newcommand{\nummstar}{13,617}
\newcommand{\numwstar}{1,648}
\newcommand{\numspec}{1,644,191}
\newcommand{\numobs}{7,389}
\newcommand{\numunique}{2,082}

\newcommand{\numsdssgal}{1,028}
\newcommand{\numsdssstar}{746}
\newcommand{\numsdssspec}{4,616}

\newcommand{\virus}{VIRUS}
\newcommand{\remedy}{\texttt{Remedy}}
\newcommand{\hetdex}{HETDEX}

\newcommand{\sdss}{SDSS}

\newcommand{\redchi}{${\chi}_{\nu}^2$}

\accepted{March 18, 2024}
\published{May 1, 2024}

\shorttitle{VIRUS Continuum Catalog}
\shortauthors{Zeimann et al.}
\graphicspath{{./}{figures/}}

\begin{document}

\title{The Hobby Eberly Telescope VIRUS Parallel Survey (HETVIPS)\footnote{Based on observations obtained with the Hobby-Eberly Telescope (HET), which is a joint project of the University of Texas at Austin, the Pennsylvania State University, Ludwig-Maximillians-Universit{\" a}t M{\" u}nchen, and Georg-August-Universit{\" a}t G{\" o}ttingen. The HET is named in honor of its principal benefactors, William P. Hobby and Robert E. Eberly.}}

\correspondingauthor{Greg Zeimann}
\email{gregz@astro.as.utexas.edu}

\author[0000-0003-2307-0629]{Gregory R. Zeimann}
\affiliation{Hobby Eberly Telescope, University of Texas, Austin, Austin, TX, 78712, USA}

\author{Maya H. Debski}
\affiliation{Department of Astronomy, The University of Texas at Austin, 2515 Speedway Boulevard, Austin, TX 78712, USA}

\author{Donald P. Schneider}
\affiliation{Department of Astronomy \& Astrophysics, The Pennsylvania
State University, University Park, PA 16802, USA}
\affiliation{Institute for Gravitation and the Cosmos, The Pennsylvania State University, University Park, PA 16802, USA}

\author[0000-0003-4381-5245]{William P. Bowman}
\affiliation{Department of Astronomy, Yale University, New Haven, CT 06511, USA}

\author{Niv Drory}
\affiliation{McDonald Observatory, The University of Texas at Austin, Austin, TX 78712, USA}

\author[0000-0001-6717-7685]{Gary J. Hill} 
\affiliation{McDonald Observatory, The University of Texas at Austin, Austin, TX 78712, USA}
\affiliation{Department of Astronomy, The University of Texas at Austin, 2515 Speedway Boulevard, Austin, TX 78712, USA}

\author[0000-0002-3559-5310]{Hanshin Lee}
\affiliation{McDonald Observatory, The University of Texas at Austin, Austin, TX 78712, USA}

\author{Phillip MacQueen}
\affiliation{McDonald Observatory, The University of Texas at Austin, Austin, TX 78712, USA}

\author{Matthew Shetrone}
\affiliation{University of California Observatories, University of California Santa Cruz, Santa Cruz, CA 95064, USA}

\begin{abstract}

The Hobby-Eberly Telescope (HET) VIRUS Parallel Survey (HETVIPS) is a blind spectroscopic program that sparsely covers approximately two-thirds of the celestial sphere and consists of roughly \nfibers\ fiber spectra. The spectra were taken in parallel mode with the Visible Integral-field Replicable Unit Spectrograph (VIRUS) instrument when the HET was observing a primary target with other HET facility instruments.  VIRUS can simultaneously obtain approximately 35,000 spectra covering 3470~\AA\ to 5540~\AA\ at a spectral resolution of $\approx$~800.  Although the vast majority of these spectra cover blank sky, we used the Pan-STARRS1 Data Release 2 Stacked Catalog to identify objects encompassed in the HETVIPS pointings and extract their spectra.  This paper presents the first HETVIPS data release, containing  \numobj\ flux-calibrated spectra obtained through 31~March~2023, as well as a description of the data processing technique.  Each of the object spectra were classified, resulting in a catalog of \numgal\ galaxies, \numqso\ quasars, \numstars\ stars, and \numunknown\ unknown sources. 

\end{abstract}



\section{Introduction} \label{sec:intro}
Over the past two decades, advances in instrumentation and in data processing power have enabled systematic surveys that have produced datasets of unprecedented size for the astronomical community.  The first digital ``megasurvey", the Sloan Digital Sky Survey (SDSS; \citealt{york2000}), has produced over a dozen public data releases containing deep $ugriz$ photometry over a quarter of the celestial sphere and optical/near-infrared spectroscopy of millions of sources.  Other examples are the Kepler mission (\citealt{kepler}), with its high-precision photometric light curves of tens of thousands of stars, and the Gaia mission (\citealt{gaia}), with its stunning compilation of precise proper motions of on order a billion stars. In a few years, the Large Synoptic Survey Telscope (LSST; \citealt{lsst}) will begin operations, producing sub-arcsecond sampling of a large fraction of the sky in multiple filters on the timescales of a few days.

In this paper we describe a new large astronomical program, the Hobby-Eberly Telescope VIRUS Parallel Survey (HETVIPS), that will be generating data regularly over the next several years, and present its first public data release.  The observations were obtained on the Hobby-Eberly Telescope (HET; \citealt{lwr98,hill2021}) with the Visible Integral-field Replicable Unit Spectrograph 
(VIRUS\footnote{VIRUS is a joint project of the University of Texas at Austin,Leibniz-Institut f\"ur Astrophysik Potsdam (AIP), Texas A\&M University (TAMU), Max-Planck-Institut f\"ur Extraterrestrische Physik (MPE), Ludwig-Maximilians-Universit\"at Muenchen, Pennsylvania State University, Institut f\"ur Astrophysik G\"ottingen, University of Oxford,
and the Max-Planck-Institut f\"ur Astrophysik (MPA). In addition to
Institutional support, VIRUS was partially funded by the National Science Foundation, the State of Texas, and generous support from private individuals and foundations.};
\citealp{hill2021}), a fiber-fed, multi-spectrograph instrument that can obtain low-resolution optical spectra of nearly 35,000 sky positions simultaneously.  VIRUS is the instrument employed in the Hobby-Eberly Telescope 
Dark-Energy Project (HETDEX, \citealp{Hill2008,gebhardt2021}).

In addition to targeted observations like HETDEX, VIRUS is operated in a parallel mode during HET observations using other facility instruments.  The parallel VIRUS data are, in essence, taken on areas of  ``blank" sky that are within approximately 10$'$ of the primary target of the observation.  The vast majority of the spectra obtained in parallel mode are simply sky observations.  Using a deep imaging catalog from the PanSTARRS project (\citealt{cham2016}), we locate objects that have fallen in the VIRUS parallel fields, and extract their spectra.

This first data release (HETVIPS DR1) is based upon the VIRUS parallel data taken between 1 January 2019 to 31 March 2023.  The raw database consists of approximately \nfibers\ spectra obtained in \numobs\ observations.  The HETVIPS DR1 Continuum Catalog consists of \numobj\ unique object spectra located over a wide region on the celestial sphere.  The spectra are classified into four groups (star, galaxy, quasar, or unknown).  The information contained in the HETVIPS DR1 Catalog, including spectral classifications and redshifts, can be used to address a wide variety of scientific projects ranging from the chemical history of the Galaxy, individual stellar systems such white dwarfs, and properties of quasars out to redshifts of 3.5.

The VIRUS parallel observations, including the sky coverage, exposure time distribution, and the spectral depth are described in \S\ref{sec:observations}.  The procedure for extracting the calibrated fiber spectra, and the extraction of the object data, are presented in \S\ref{sec:remedy} and \S\ref{sec:objextract}, respectively.  The continuum catalog construction is reported in \S\ref{sec:catalog}. The spectral classification methodology is reviewed in \S\ref{sec:diagnose}.  The HETVIPS DR1 catalog data products and the data distribution system for the spectra are presented in \S\ref{sec:datadist}, and a brief summary of HETVIPS is provided in \S\ref{sec:summary}.

\section{Observation Technique}
\label{sec:observations}

\subsection{Instrumentation}

The HET is a fixed-altitude (zenith angle of~$35^{\circ}$) telescope with an 11-m diameter mirror composed of 91 identical hexagonal spherical mirror segments.  During an observation, the telescope superstructure remains stationary; the tracking is provided by a platform located at the prime focus that carries the wide field corrector optics and fiber optic inputs (see \citealp{hill2021}). The telescope can access all points on the sky \hbox{at $-10.3^{\circ} < \delta < +71.6^{\circ}$;} the maximum length of the tracks is dependent upon declination, with the available track time generally increasing from approximately one hour at the southern limit to 2.5 hours for the northernmost observations.

The HET operates in a queue mode (\citealt{shetrone2007}) and has three facility instruments (all fiber fed): the aforementioned VIRUS, the Low-resolution Spectrograph~2 (LRS2; \citealt{chonis16}), an optical/near-IR, $R \approx 2000$ integral field spectrograph, and the Habitable Zone Planet Finder (HPF; \citealt{mahadevan18}), a precision radial velocity spectrograph that operates in the near-infrared. The input for VIRUS consists of 78 integral field units (IFUs) positioned in a grid that nearly spans the HET's 22$'$ diameter corrected field of view, positioned on an input head mount plate.  The fibers for the LRS2 and HPF are located in the central region of the plate (see Figure 6 of \citealp{hill2021}). The fiber inputs for all three instruments share a common focal surface and shutter.

All of the HETVIPS observations were obtained with VIRUS.  This instrument is comprised of a set of identical 78 integral field spectrographs that produce $\approx$~35,000 spectra covering the wavelength range 3470\AA\ -- 5540\AA\ at a resolving power $R \approx 800$. The instrument's sky footprint is 56 \sqam\ within an 18\arcmin\ diameter field covered by the IFU grid layout; the fill factor is $\simeq$ 1/4.5.  Each IFU of 448 fibers covers a solid angle of approximately 50\arcsec\ x 50\arcsec\ and  feeds a pair of spectrographs.
The separation between IFUs is 100\arcsec.
The individual fibers are 1.5\arcsec\ in diameter,
arrayed in a hexagonal pattern with spacing between the fiber centers of 2.2\arcsec.  During HETDEX observations, a triangular dither pattern of three exposures nearly fills these gaps ($\sim$94\% sky coverage).  All of the VIRUS parallel observations are un-dithered; thus, the sky coverage within an IFU is roughly one-third. A good analogy for the HETVIPS observations would be an imaging instrument that covered an 18\arcmin\ diameter field of view, but only slightly more than 1 in 14 pixels of the device record data (1/3 factor for un-dithered observations, 1/4.5 factor for the IFU coverage).  

\subsection{Observations}

If the HET is acquiring an observation where the primary instrument is not VIRUS and the exposure time is greater than five minutes, a VIRUS exposure is obtained. The exposure time is set so that the VIRUS observation does not impact the primary observation (i.e., VIRUS is employed in ``parallel'' mode). The primary targets (using the LRS2 or HPF) of the VIRUS parallel observations are located near the center of the field of view, and consist of a wide range of classes of objects; thus, the HETVIPS survey is not contiguous, but consists of a collection of a large number of pointings throughout the HET observable sky ($\sim$2/3 of the total 4$\pi$ sky).  Also, the VIRUS IFUs have a $\sim$1/3 fill factor with the intent that a three-point dither pattern results in a near full sky coverage for a given IFU.  However, all of the parallel mode observations in HETVIPS are un-dithered, leaving a gap of 2.2\arcsec\ between the centers of each of the 1.5\arcsec\ diameter fibers.  While there are more than \nfibers\ fiber spectra in our dataset, the VIRUS fibers primarily collected light from the night sky and rarely covered detectable astronomical sources.

\begin{figure*}[t]
	 \includegraphics[width=2\columnwidth]{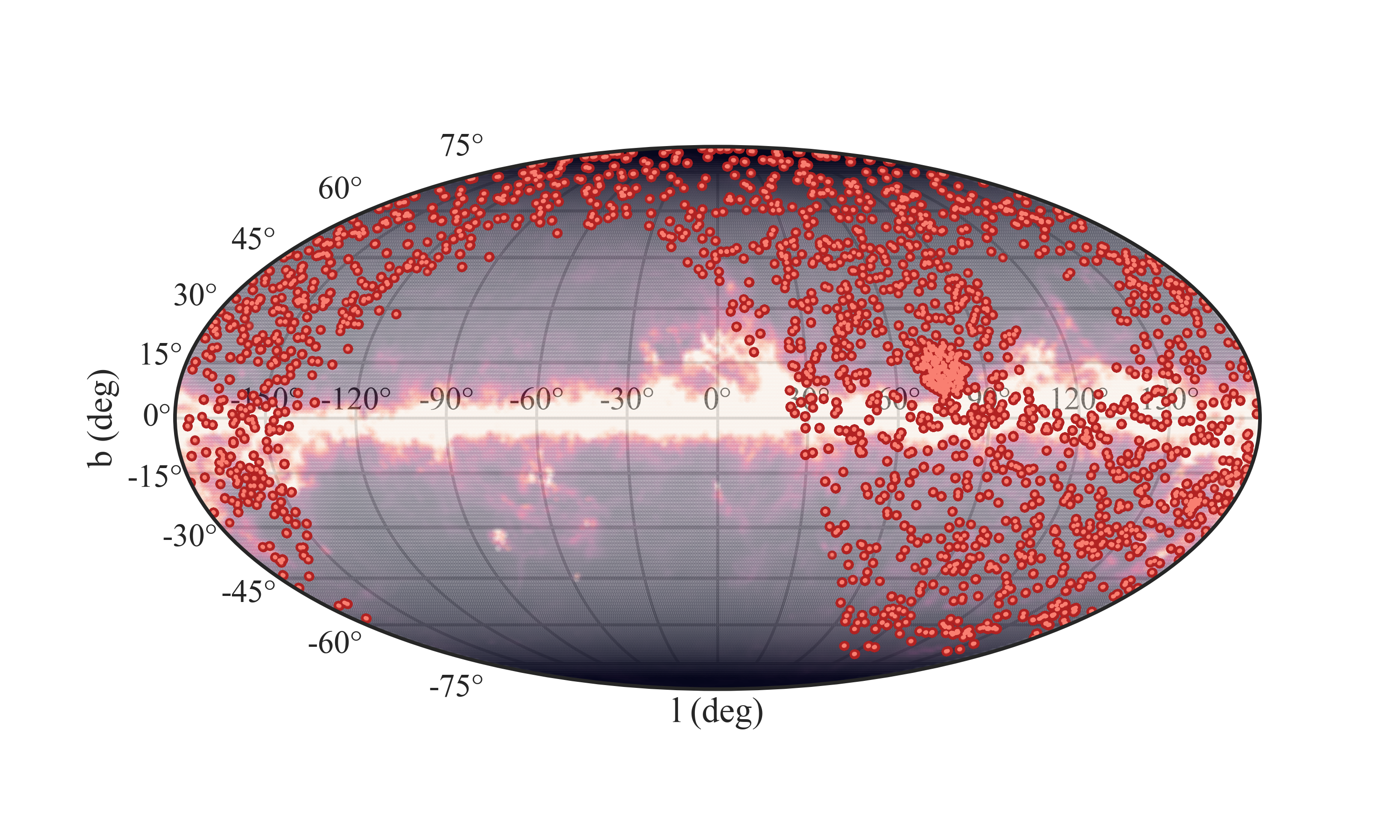}
	\caption{The location of the HETVIPS observations (red dots) displayed in Galactic coordinates.  The background is an infrared dust map of the Milky Way \citep{sfd98}. Many of the pointings have multiple exposures arising from repeated visits to a given primary target (e.g., HPF radial velocity monitoring of a bright star).  The fixed azimuth of the HET results in an observable declination constraint, producing two gaps in coverage; the small hole in the upper right \hbox{($\delta > +71.6^{\circ}$)} and the large area on the lower left \hbox{($\delta < -10.3^{\circ}$).}}
	\label{fig:sky_coverage}
\end{figure*}

A total of \numobs\ observations taken in this mode from 1 January 2019 through 31 Mar 2023 met our criteria (see S~\ref{subsec:astrometry}) for inclusion in the HETVIPS database. Figure \ref{fig:sky_coverage} displays the location of each HETVIPS pointing in Galactic coordinates. Of the \numobs\ pointings, \numunique\ are unique celestial positions.  Nearly 50 locations have more than 10 visits (primarily HPF observations). The HETVIPS observations cover both the Galactic plane and high-Galactic latitudes.  The total unique area covered by HETVIPS is roughly 26 deg$^2$.

\begin{figure}[t]
	 \includegraphics[width=1\columnwidth]{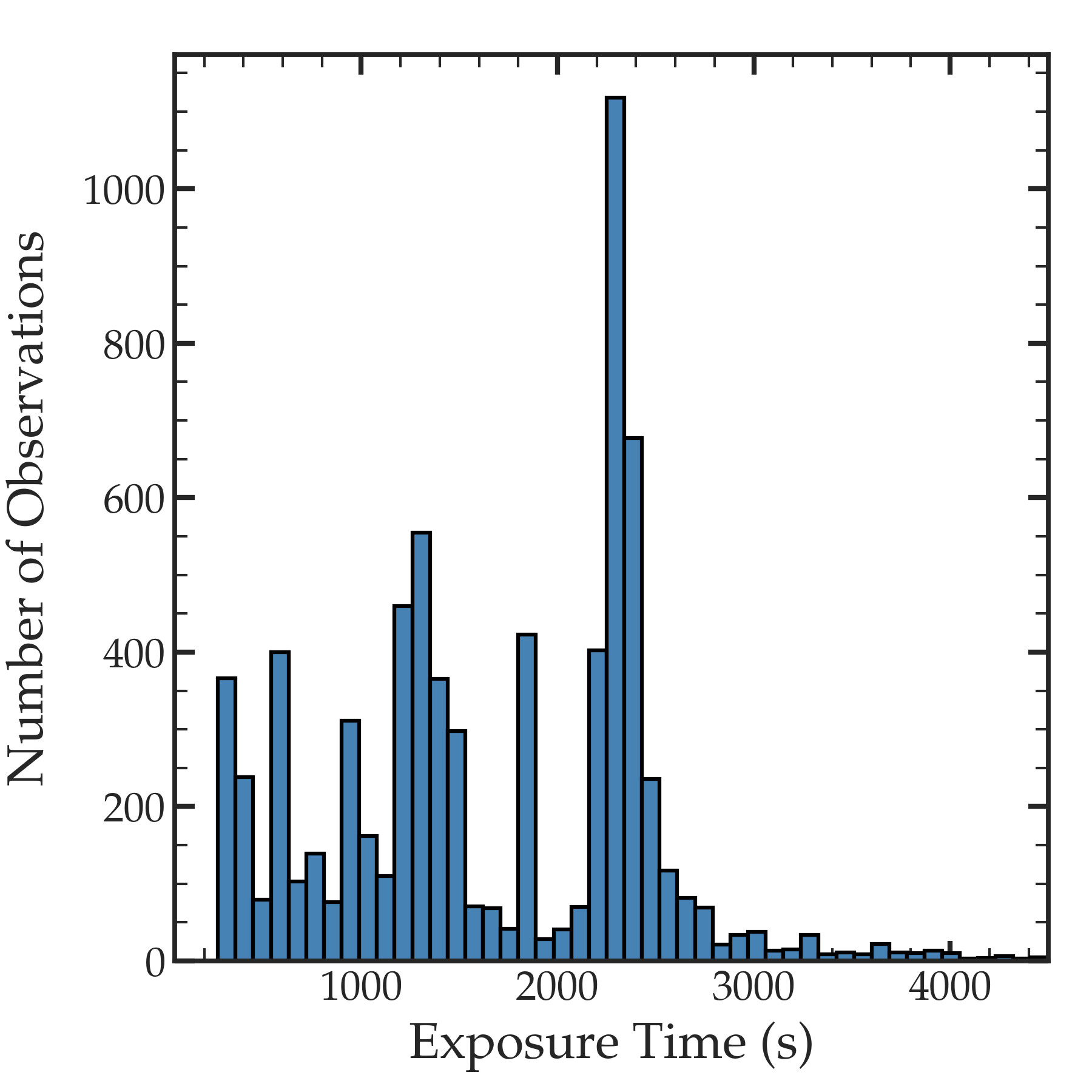}
	\caption{The histogram of exposure times for the \numobs\ pointings in HETVIPS DR1.  HET observations shorter than 300~s are excluded from HETVIPS.  The maximum exposure time is limited by the travel constrains of the HET's prime focus tracking system.}
	\label{fig:exptime}
\end{figure}

Figure \ref{fig:exptime} presents the histogram of the exposure times for the HETVIPS pointings.  The exposure times range from 300 - 4500 seconds; the most common exposure times are multiples of 300 seconds, and the median exposure time is 1800 seconds. 

Given the blind nature of the survey, and the sparse sampling of the instrument, a catalog of sources in the HETVIPS fields is required to provide astrometric and photometric calibration of the data.    Such a catalog should contain photometric measurements in the VIRUS bandpass, have high astrometric precision,  and reach relatively faint magnitudes.  The Pan-STARRS  (\citealt{cham2016}) Data Release 2 Stacked Object Catalog (PS1 DR2, \citealt{flew2020}) provides an excellent match to our requirements and fully encompasses the HETVIPS observations.

\section{Processing the Fiber Spectra}
\label{sec:remedy}
The \virus\ parallel observations were processed by \remedy\footnote{https://github.com/grzeimann/Remedy}, a software package designed to produce flux-calibrated spectra for every VIRUS fiber, as well as create spectra for objects of known position in the  \virus\ footprint.  \virus, at full capacity, is composed of 78 IFUs, with each IFU feeding two spectrographs. Each of these spectrographs has its own CCD, and each CCD is read through two amplifiers.  As a result, VIRUS data contains 312 distinct sets of CCD parameters, each with its own bias structure, gain, read noise, and dark current.  A full set of VIRUS calibration observations consists of six types of data frames:  science, twilight, flat-field from a Laser Driven Light Source (LDLS), arc lamp, bias, and dark.  

\subsection{Overscan Subtraction and Construction of Master Frames}
\label{subsec:ovescan}
The data from each of the 312 amplifiers has an overscan region.  After excluding the first two columns of the overscan due to potential ``bleeding'' from the data section, the overscan's biweight location \citep{Beers1990} is used to create a single value that is subtracted from the image.  The overscan region is trimmed from the frame, leaving only the data section. 

To perform bias and dark current subtraction, distortion mapping, wavelength calibration, fiber normalization, and poor data identification, we create master frames for each of the six types of VIRUS exposures (bias, dark, arc lamp, twilight, LDLS, and science).  For bias, dark (all dark exposures are 360s), twilight, and LDLS exposures, we collect all observations in a calendar month and perform a median average.  Combining the data produces a master frame for each calibration class for all 312 amplifiers.   The Hg and Cd lamp frames are combined separately and these ``lamp masters" are used to produce a master arc frame. For all of the master frame creations, we ignore amplifier exposures that were all zeros;  this situation occurs rarely and is due to electronic failures.  

With 312 different sets of electronics, VIRUS is a complicated instrument.  As a result, a given observation, night, or longer period can suffer from a variety of problems.  To broadly identify any issue, we use our median master science frames, which are constructed from all \hetdex\ observations taken in a calendar month ($\sim50-200$).  This median-filtered image should be devoid of real sources, containing only signal from an ``average sky''. We construct a pseudo-error image for this master frame that mimics the error of a single observation, rather than that of the stacked master frame.  This approach allows identification of issues in the sky-subtracted master frame spectra that can produce systematic variations beyond the threshold of the typical statistical error (i.e., when the systematic error is greater than this threshold times the average statistical error).  For example, charge traps are not well-modeled in our fiber normalization process and produce systematic outliers in sky-subtracted master science frame spectra.  

\subsection{Bias and Dark Subtraction}
\label{subsec:bias}
There are two classes of bias structure in a VIRUS amplifier. The first is a slowly-evolving large-scale structure whose amplitude is typically much less than the amplifier read noise, and is easily removed from a bias master frame.  The second type of structure occurs on a pixel-to-pixel scale, and resembles an interference pattern.  This feature changes rapidly throughout the night and is not easily eliminated.  It is, however, only $\lesssim$1 electron in amplitude, and even less when averaged over a fiber profile in the trace direction.  We use the master bias frames to subtract the large-scale structure as well as any coherent interference pattern.

The dark current in VIRUS amplifiers produce $\lesssim$0.5 electrons in a 360s exposure.  The VIRUS amplifiers are subject to spurious charge on the side of the detectors closest to the amplifier readouts.  The spurious charge can be $\sim$3-4 electrons (in 10\% of the amplifiers) and decays over the course of approximately 100 rows.  These features are locked into the master dark frames and scaled by exposure time for subtraction.

\subsection{Gain Correction}
\label{subsec:gain}
The output of the amplifier images are in ADUs (analog-to-digital units).  These values to are converted to electrons using the gain value for the amplifier in the image header.  This gain was calculated in the laboratory prior to the detector's installation in the spectrograph.

\subsection{Error Propagation}
\label{subsec:errorprop}
The error value for each pixel is propagated from the read noise recorded in the frame header and the Poisson error from the number of electrons detected.  No additional correction is made in the low-count regime, where the error inferred from the observed counts differs from that derived from the true number \citep{Gehrels1986, Kraft1991}.  It is extremely rare for any of the pixels that contribute to our spectral extractions to be in this low-count regime due to contribution of sky continuum/emission illuminating the fibers.  

\subsection{Pixel Mask}
\label{subsec:pixelmask}
The master dark frame is used to identify ``hot" pixels and low-level charge traps which must be masked in subsequent processing.  The median value in each column is subtracted from the master dark to remove the large-scale columnar pattern, then the median value of each row is subtracted to similarly eliminate low-level patterns in the row direction.  In the final step, a $\sigma$-clipping algorithm identifies and masks all pixels that are $5\,\sigma$ outliers.  The vast majority of these outliers are either hot pixels or located at the base of low-level charge traps.

\subsection{Fiber Trace}
\label{subsec:trace}
The traces of the individual fibers on the detectors are identified from high signal-to-noise ratio LDLS exposures.  For each wavelength pixel for an individual fiber, the peak pixel and its two neighboring pixels are used to define a quadratic function.  The peak of the fitted quadratic function defines the center of the trace of the fiber for a given column.  A third-order polynomial fit to the trace centers is used to produce a smooth curve for the fiber.

To monitor the behavior of the fiber trace, the trace map of each frame is compared to the trace from the current master LDLS frame.  Over the two years of HETVIPS observations the system has remained quite stable.  The night-to-night drift in position, driven primarily by changes in the ambient temperature, is $\lesssim 0.2$ pixels for an individual fiber; the traces appear to move around a stable location and do not experience a secular drift.

\subsection{Fiber Wavelength Calibration}
\label{subsec:wavelength}
The wavelength calibration for each fiber is obtained from the Hg and Cd arc lamp master frame exposures.  The centers of the lines are identified with a peak-finding algorithm with a priori knowledge of the strong line wavelengths. A third-order polynomial fit to the line centers creates a stable wavelength solution, and, as was the case for the fiber trace, typical shifts of $\lesssim$0.2 pixels occur due to changes in the ambient temperature.  The wavelength solutions have typical rms errors of~0.05~\AA.  There are rare failures in the wavelength solution process ($\approx$~1\% of the time), which lead to a full masking of the amplifier.

\subsection{Spectrograph Scattered Light}
\label{subsec:scatteredlight}
Low-level wings of light, scattered by residual imperfections in the surface finish of the VIRUS spectrograph optics, are evident on the point spread functions of the fiber images on the detectors \citep{hill2021}. These wings are
modeled as a power-law, in which the monochromatic intensity of a given fiber has a profile that falls off inversely with distance from the center of the trace.  
When the sum of the signal from the central cores of all the fibers is compared to the total light recorded by the detector, it appears that approximately 3-4\% of the incident fiber light is scattered out of the cores of the point-spread functions into a relatively smooth background.
This value of total scattered light agrees with direct measurements on isolated fibers (\citealt{hill2021}, Fig. 11) and with expectations for the surface finish specifications of the optics. 
The signal in the fiber gaps is used to measure the background light and interpolate this background across an amplifier.  This background model is subtracted from the data.  We consider the light as lost from the system and do not attempt to correct for this loss of signal in post-processing of the data.

\subsection{Spectral Extraction}
\label{subsec:extraction}
The fiber spectra are extracted by summing the central five pixels, i.e., $\pm 2.5$ pixels about the trace.  The two outer pixels in this range are assigned linear weights which represent the fraction of the pixel that is within 2.5 pixels of the trace.  A normalized $\chi^2$ for each fiber is calculated by comparing the fiber profile to that in the master LDLS frame.  This $\chi^2$ provides a metric by which to flag cosmic rays and bad columns that fail to follow the expected profile shape.  We propagate the error for the spectrum in quadrature, using the appropriate weight for the outer pixels. When the spectrum is extracted, linear interpolation is used to place it and the associated propagated errors on a rectified linear wavelength grid between 3470~\AA\ and 5540~\AA\ at a dispersion of 2~\AA\ per pixel.

\subsection{Fiber Normalization}
\label{subsec:normalization}
The fiber normalization corrects the signal in each fiber relative to the average throughput.  We utilize two master frames to achieve this goal: the twilight and LDLS.  Since we did not correct our CCDs for pixel-to-pixel variations, our summed extracted spectra still include the imprint of small scale quantum efficiency differences.  The LDLS intrinsic spectrum is smooth with no sharp features.  We use the LDLS master frame to extract spectra following \S\ref{subsec:extraction}, and normalize our extracted spectra using a biweight smoothing spline (kernel$\sim$80~\AA).  The resulting product provides a correction for our small wavelength scale fiber normalizations.  

The master twilight frame is employed to evaluate the larger wavelength scale fiber normalizations by making the assumption that, averaged over many twilight frames, the illumination across our $18\arcmin$ field-of-view is uniform. We again extract master twilight frame spectra following \S\ref{subsec:extraction}, and fit a biweight smoothing spline (kernel$\sim$80~\AA) for the large-scale normalization.
We divide all science fiber spectra by the appropriate normalization and propagate this division into the error spectra.

\subsection{Spectral Masking}
\label{subsec:masking}
A VIRUS observation of one exposure contains up to 78 IFUs, each with 448 fibers, i.e., an individual exposure can contain up to 34,944 separate spectra.  As each IFU has two CCDs and each CCD has two amplifiers, there are a number of CCD defects which must be identified and flagged.  The pixel masking procedure in the raw data was presented in \S\ref{subsec:pixelmask}.  During the fiber extraction, when collecting the relevant pixels for a given column and fiber, if just one of the five pixels over which we sum to extract the spectra (see \S~\ref{subsec:extraction}) has a masked value, we conservatively exclude the spectrum at that column/wavelength. Similarly, when evaluating the normalized $\chi^2$ for each wavelength of each fiber's spectrum (see \S\ref{subsec:extraction}), if the normalized $\chi^2$ is greater than five, that spectral wavelength is masked, as well as columns $\pm 2$ pixels from the high $\chi^2$ column.  Since the primary cause for high $\chi^2$ values is the presence of a cosmic ray, our conservative masking approach avoids the contribution from the fainter wings of these artifacts.  Finally, we also exclude fibers with normalizations less than 50\% (i.e., low throughput or dead fibers).  

As discussed in \S\ref{subsec:ovescan}, we construct master science frames from all \hetdex\ observations in a calendar month.  We use the fiber normalizations to build a single sky spectrum for our master science images, subtract that sky from our master science spectral extractions, and construct a pseudo-error image for this master frame that mimics the error of a single observation.  Data are flagged for removal if they are less than $-0.5$ or greater than 0.75 times the pseudo statistical error.   

Finally, an entire fiber is removed if more than 20\% of the columns in the fiber are flagged.  An entire column of an amplifier is eliminated if the fraction of flagged pixels or the fraction of removed spectra exceeds 20\%.

\subsection{Sky Subtraction}
\label{subsec:skysubtract}
We construct a single sky spectrum for each exposure using the biweight location at each wavelength, excluding fibers whose average emission is 2-$\sigma$ above the background.  This single composite sky spectrum is subtracted from each fiber.  After removing this sky model, the ``blank'' sky fibers are used to create a smoother and more local sky residual employing a Gaussian kernel in the fiber and wavelength direction.  The kernel has a sigma of seven fibers in the fiber direction and 14~\AA\ in the wavelength direction.  As expected, this approach to sky subtraction performs best in observations with a large fraction of ``sky'' fibers and encounters challenges when the field of an IFU contains extended objects with angular sizes comparable to the IFU, or large numbers of unresolved sources (e.g., globular clusters).

\subsection{Astrometric Calibration}
\label{subsec:astrometry}
The astrometric calibration for each of the \nfibers\ fibers is provided by a comparison to the sources in PS1 DR2.  We proceed IFU by IFU, collapsing each sky-subtracted spectrum into a single value (integrated over the PS1 $g$ filter), interpolating this signal onto a uniform grid on the sky using the focal plane information, detecting point sources in the collapsed image, and matching these detections to the objects in the PS-1 DR2 catalog.  We then fit for a shift and rotation from the initial VIRUS astrometry provided in the data frame information.  The typical positional shift is less than 2\arcsec\ and the rotation is less than $0.1^{\circ}$.  The biweight rms deviation of individual stars' positions about the solution for 95\% of our fields is $\lesssim 0.5\arcsec$.  We define the astrometric error for a field centroid as the individual star rms divided by the square root of the number of stars used for the solution.  The astrometric error for 95\% of our field centroids is $\lesssim 0.15\arcsec$.  For a VIRUS parallel observation to be included into the final HETVIPS catalog, we required that at least 5 stars were used to get the astrometric solution, that the astrometric error for the field centroid was $\lesssim 0.25\arcsec$, and that the biweieght rms deviation of the individual positions was $\lesssim 0.5\arcsec$. 

In addition to the astrometric error of the field there are also uncertainties in the focal surface position angle, IFU seat positions, IFU seat rotations, and the fiber positions within an IFU.  As previously noted, the focal surface position angle is initially provided by the data frame headers and the re-fitting for the angle involves a shift of less than $0.1^{\circ}$.  Assuming an accuracy of half of the adjustment, $0.05^{\circ}$ at a 4\arcmin\ distance from the field center is a positional uncertainty of 0.2\arcsec.  The IFU seat positions are known to 0.05\arcsec\ accuracy from lab measurements and confirmed on-sky (Gebhardt et al. 2021).  The IFU seat rotations are known to $0.2^{\circ}$, which amounts to a 0.08\arcsec\ error at 22\arcsec\ from the IFU center.  Finally, the fiber positions within the IFU are known to 0.05\arcsec\ from laboratory measurements (\citealp{hill2021}).  Taking all of these astrometry errors in quadrature, including the field centroid error and the initial error in the PS1 catalog, there is a net expected astrometric error of 0.30\arcsec\ for fiber positions. 

\subsection{Flux Calibration}
\label{subsec:fluxing}

\begin{figure}[t]
	 \includegraphics[width=1\columnwidth]{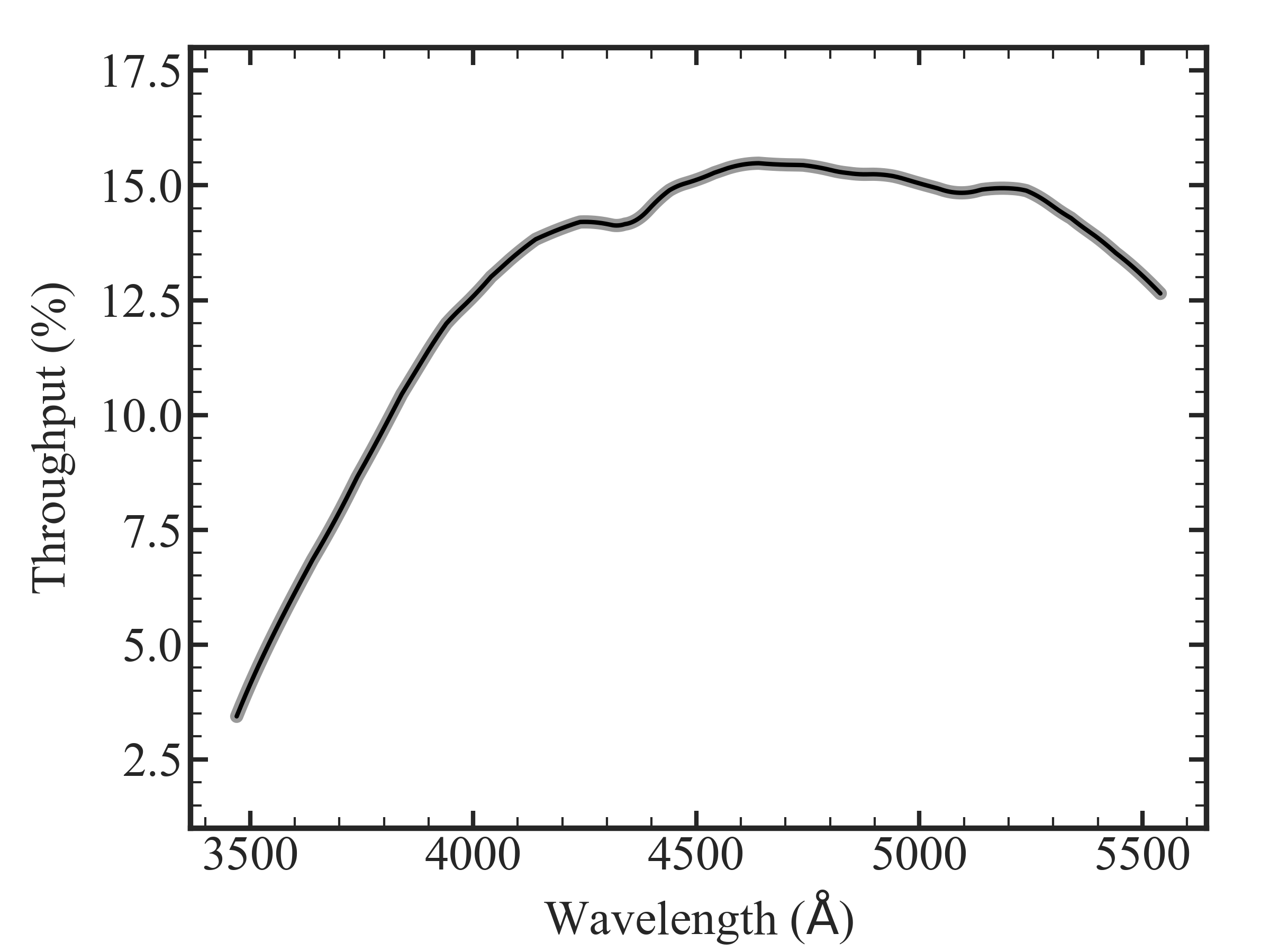}
	\caption{The \remedy\ VIRUS throughput curve used for the HETVIPS flux calibration.  The curve is quite similar to that of Figure 17 in \citet{hill2021} despite a different derivation and this work's throughput curve is plotted for an effective mirror illumination of 50 m$^2$ rather than a full 10-m aperture.}
	\label{fig:throughput}
\end{figure}

During an observation the HET guide cameras (GCs) are constantly providing measurements of the system's $g$-band throughput.  The \remedy\ VIRUS throughput curve (see Figure \ref{fig:throughput}) for an effective 50 m$^2$ mirror illumination is adjusted based on this GC information.  During an observation the HET tracker moves across the primary mirror; thus the illumination of the primary mirror is constantly changing during an observation.  We use the GCs to monitor the tracker position for each exposure and model the integrated mirror illumination using HETILLUM.  After correcting for the GCs' throughput and the primary mirror illumination, we use the set of PS1 DR2 sources from the astrometry calibration, create their spectra (see \S\ref{subsec:pixelmask}), construct a synthetic PS1 $g$ magnitude, and correct any average residual offset between the synthetic and catalog magnitudes.  The residual correction is usually less than 10\%.

\begin{figure}[t]
	 \includegraphics[width=1\columnwidth]{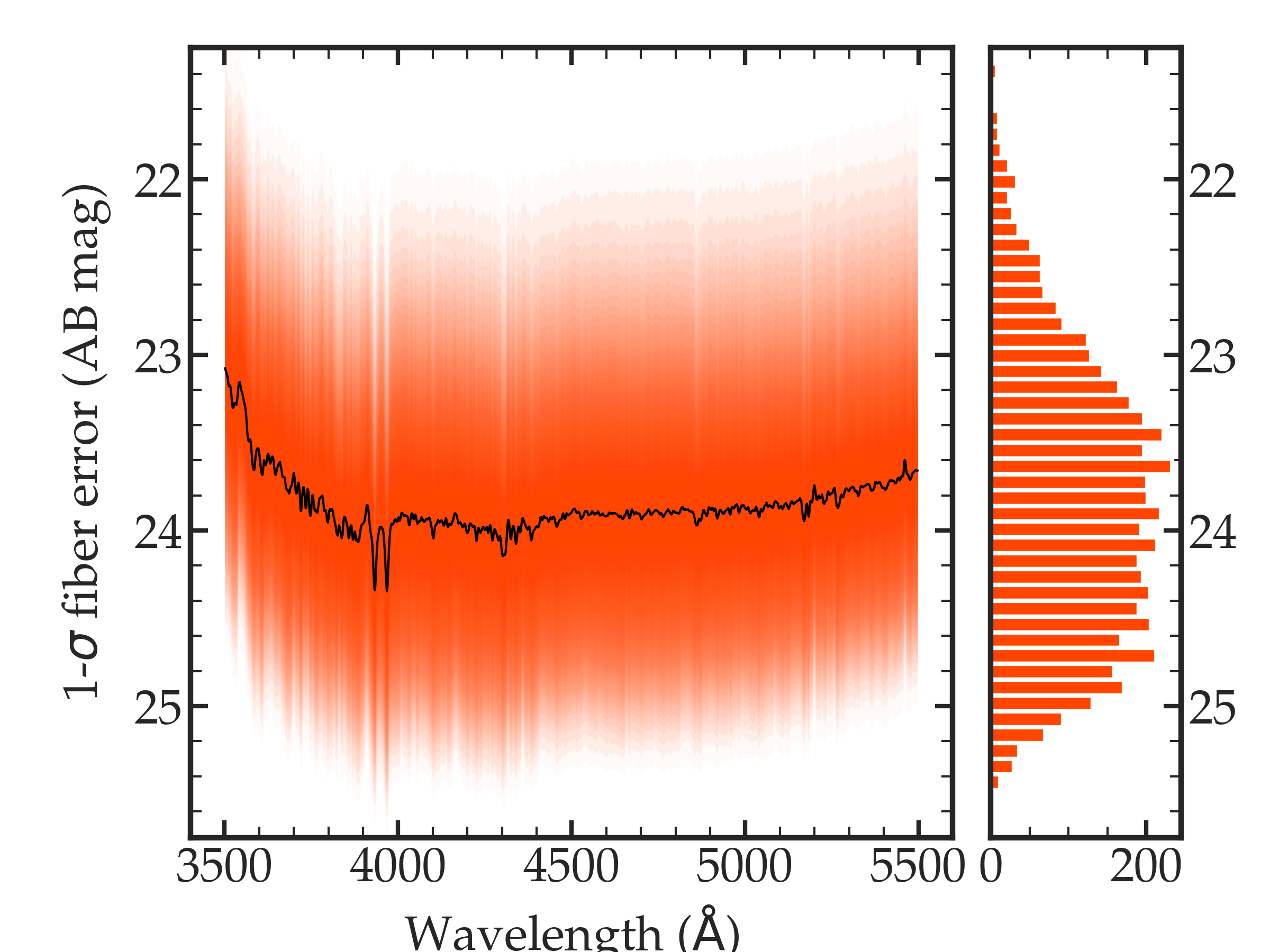}
	\caption{The left panel displays the median 1-$\sigma$ fiber error spectrum in the sky per resolution element (spectral and spatial, in orange) for the HETVIPS observations (essentially the background uncertainty).  The average for all \numobs\ exposures is shown in black.  The histogram of the AB magnitude sensitivities at 5000~\AA\ is shown on the right.}
	\label{fig:sensitivity}
\end{figure}

 The median-average flux-calibrated fiber error spectrum is calculated for each exposure; given that the vast majority of the spectra are of the sky, this quantity is, in essence, the 1-$\sigma$ noise in the background. The distribution of these 1-$\sigma$ error arrays are shown in Figure \ref{fig:sensitivity}.  The 5$^{\rm th}$--95$^{\rm th}$ percentiles span three photometric magnitudes at all wavelengths, with the greatest sensitivity at the night sky Ca~II H and K absorption features.  The large span in sensitivity is due to the parallel nature of the observations.  HPF is a near-IR instrument that is primarily used during bright moonlight conditions; most of the LRS2 data are acquired when the sky brightness is considerably lower.  In addition, the HETVIPS observations have a large range of exposure times.  These two factors result in a sizeable spread in sensitivity of the exposures.

The 1-$\sigma$ of an individual fiber can be roughly converted to a 1-$\sigma$ continuum depth for a point source using a factor of five.  Details about source extraction and fiber covering fraction for HETVIPS can be found in \S\ref{sec:objextract}.  As an example, if the 1-$\sigma$ of an individual fiber is 24.0 AB magnitude at 4500~\AA\ (which is near the average of the HETVIPS sensitivity), then the 1-$\sigma$ continuum depth of a point source for that observation and wavelength would be $\approx$22.3 AB.

\section{Extraction of the Object Spectra}
\label{sec:objextract}

\begin{figure*}[t]
	 \includegraphics[width=2\columnwidth]{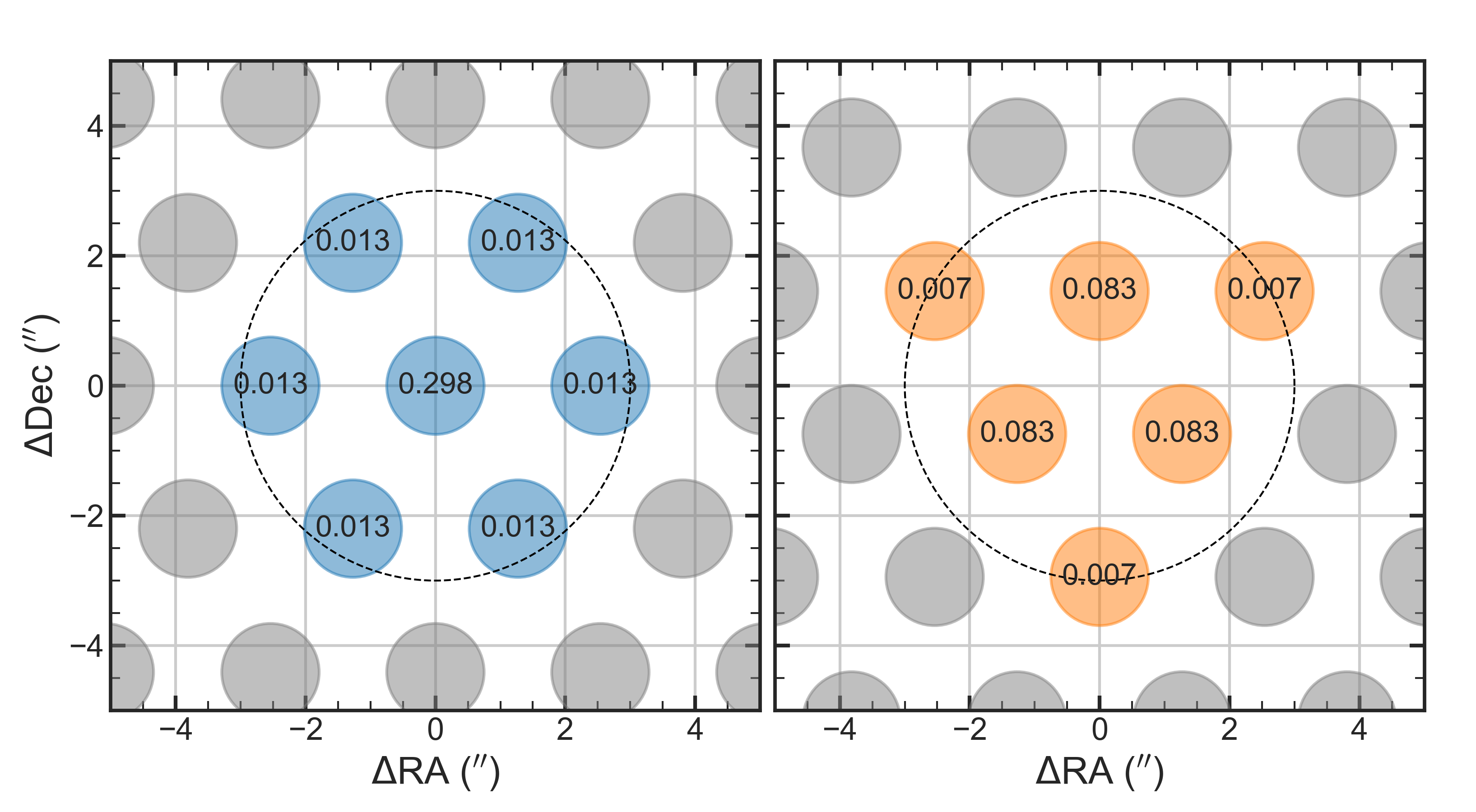}
	\caption{The filled circles represent fiber footprint on the sky illustrating the 1.5\arcsec\ diameter.  The spacing between the fiber centers is is 2.2\arcsec ; this expanded view of an IFU shows the 1/3 fill factor for a given VIRUS parallel observation.  The values in each fiber are the covering fraction for a point source model centered at 0,0 with 1.8\arcsec\ FWHM.  The covering fraction is the fraction of light from a source that the fiber is collecting.  This covering fraction changes with wavelength due to differential atmospheric refraction.  The dashed line is a 3\arcsec\ aperture and is the radius at which we collect fibers for extraction.  The left and right panels show the two possible extreme source locations with respect to the VIRUS fibers: a source centered on a fiber (left) and a source that fell between the fibers (right).}
	\label{fig:fiber_coverage}
\end{figure*}

Given HETVIP's sparse and rather coarse sampling (compared to the seeing disk) of the sky, it is not a straightforward task to reconstruct an object's spectrum from VIRUS parallel observations.  Complicating matters further, differential atmospheric refraction (DAR) shifts an objects location with respect to the VIRUS fibers by $\sim$1\arcsec\ across 3470~\AA\ - 5540~\AA. (The variation of the effects of differential refraction between exposures is mitigated by the fact that the HET operates at a fixed zenith angle.)  Even if a source falls directly on a fiber, the fraction of the object's light collected by the fiber changes as a function of wavelength.  Additionally, since the average seeing conditions for HETVIPS is 1.8\arcsec, the fraction of light collected by the closest VIRUS fiber is typically $\lesssim$30\% at any given wavelength.  However, if we can estimate the fraction of light each fiber collects within a reasonable radius, we can convert the observed fiber spectra into a single object spectrum.

\begin{figure*}[t]
	 \includegraphics[width=2\columnwidth]{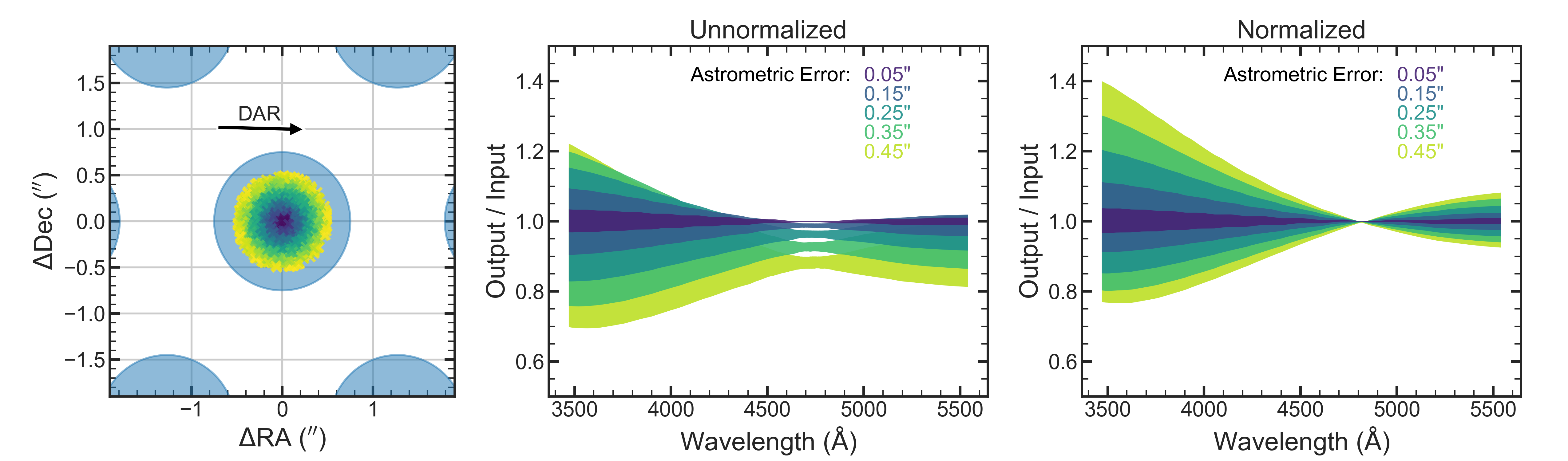}
	 \includegraphics[width=2\columnwidth]{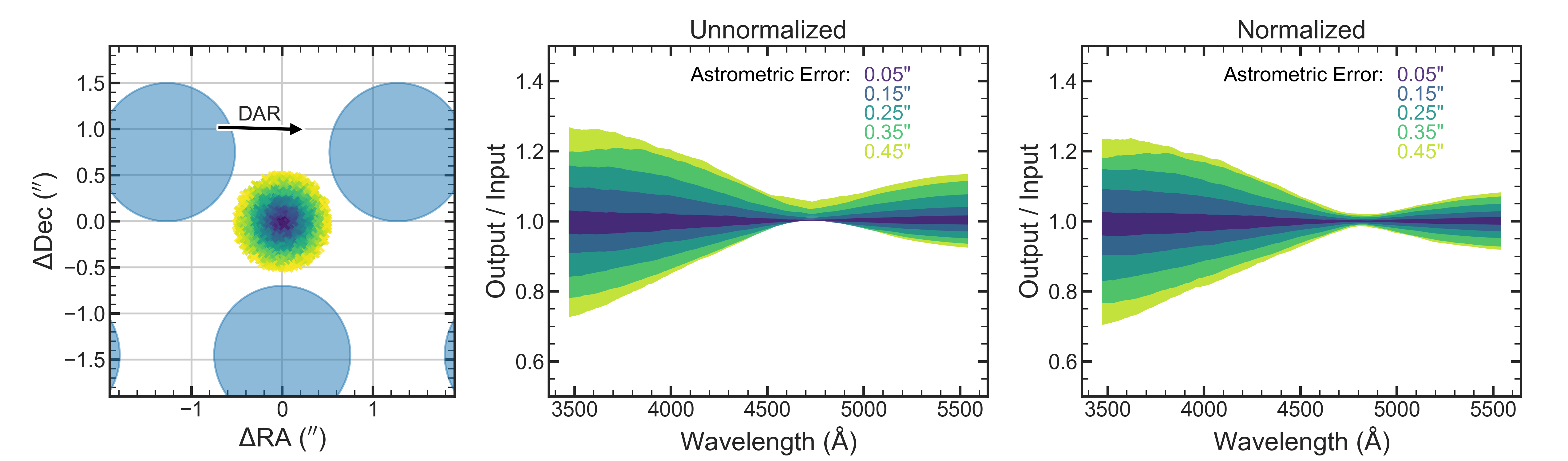}
	\caption{Simulations of 2000 sources for two different locations with respect to the VIRUS fibers were created to determine the effect of astrometric uncertainty on extracted object spectra.  Left Panel: The blue circles represent VIRUS fibers in IFU coordinates. The 'X' markers are colored by radial distance and represent the 2000 simulated source input positions.  The assumed location of the source is (0,0) and the seeing has a FWHM of 1.8\arcsec.  A black vector arrow shows the direction and magnitude of DAR across the VIRUS wavelength coverage; this change is the net effect of a source moving in the IFU plane.  The effect is applied equally for the simulated sources and the extracted output objects.  The two rows show the simulation for two scenarios: astrometric offsets centered on a fiber and in-between fibers.  Middle Panel: The $16^{\rm th}$-$84^{\rm th}$ percentile ranges in binned astrometric offsets for the extracted spectrum (Output) versus the simulated spectrum (Input).  The larger the offset the larger the systematic and relative flux calibration issue.  In the case of an object centered in the fiber (top row), larger offsets tend to have less extracted flux because the assumed location is centered on the fiber while the true location is offset, i.e., the true coverage fraction is lower than expected and thus and requires a larger normalization than the assumed applies.  Right Panel: The same $16^{\rm th}$-$84^{\rm th}$ percentile ranges in binned astrometric offsets for the extracted spectrum (output) versus the simulated spectrum (input), but we first normalize the output spectrum to the input spectrum.  This step, as will be seen later, is more reflective of the relative flux errors expected in the HETVIPS continuum catalog.}
	\label{fig:astrom_flux}
\end{figure*}

We approach object extraction independently, i.e., we do not attempt to extract many objects simultaneously but instead extract one object from one observation at a time.  For any given right ascension and declination, we first collect all fiber spectra within a 7\arcsec\ radius.  Using the GC measurement of the seeing, we construct a Moffat PSF model ($\beta = 3.5$, \citealt{Moffat1969}).  At every wavelength, we shift our fiber locations following our DAR models and convolve that PSF with the VIRUS fibers.  This estimates the fraction of the object's light covered by each fiber.  Figure \ref{fig:fiber_coverage} displays the extreme variations in fiber coverage of a standard point source and average seeing conditions for the HETVIPS survey~(1.8\arcsec).  We limit our extractions to only fibers that intersect with a 3\arcsec\ aperture.  Using these fiber coverage values as weights, we first normalize the weights to one, retaining the normalization value, then perform a weighted extraction using the \citet{Horne1986} optimal extraction formula.  Finally, the resultant spectrum is corrected to a total flux using the normalization value.  Within a 3\arcsec\ aperture, there are typically seven VIRUS fibers with a total fiber coverage between 30-40\%.

Because the total fiber coverage of a given source is relatively low, small absolute errors in fiber weights can be magnified.  We examined this effect by simulating 2000 sources with a known input spectrum.  We assume the source is always at the same location relative to the fiber centers and thus the fiber coverage weights for object extraction are identical,
but we perturb the simulated source input location so the true fiber coverage and fiber spectra change.  We uniformly distribute the offset location in radii between 0.0-0.5\arcsec and randomly distribute the angle.  This procedure is performed for two different configurations: the object centered on a VIRUS fiber, and the object positioned equidistant between three fibers.  The simulations and net effect on relative flux calibration can be seen in Figure~\ref{fig:astrom_flux}. For astrometric errors of 0.3\arcsec\ as calculated in \S\ref{subsec:astrometry} we expect a 20-25\% relative flux error in the blue and a 5\% flux error in the red. We also performed simulations where we varied the seeing of the PSF and found that an incorrect assumption in the seeing of 0.2\arcsec\ results in only a 7\% flux error in the blue and $<$1\% error in the red (in~1.8\arcsec\ images).  The impact of an incorrect PSF is smaller than that due to the estimated astrometric errors. Finally, errors in our DAR model are expected to be $<$0.05\arcsec\ across our wavelength range and have minimal effect on our relative flux calibration ($<$3\% in the blue).

For resolved objects, a point source model is incorrect but generally only induces a small relative flux calibration issue (typically $<$5\%) since in the bulk of the cases most of the spectral information is produced from a single fiber (see Figure \ref{fig:fiber_coverage}).  The measured flux from objects whose angular scales are considerably larger than several arcseconds will obviously be underestimated (perhaps by a large factor) by this technique, but the flux from sources whose scale is comparable to a few arcseconds can be over or underestimated, depending upon the scale size and location of the center relative to the fiber grid.

\section{HETVIPS DR1 Catalog}
\label{sec:catalog}

\begin{figure}[t]
	 \includegraphics[width=1\columnwidth]{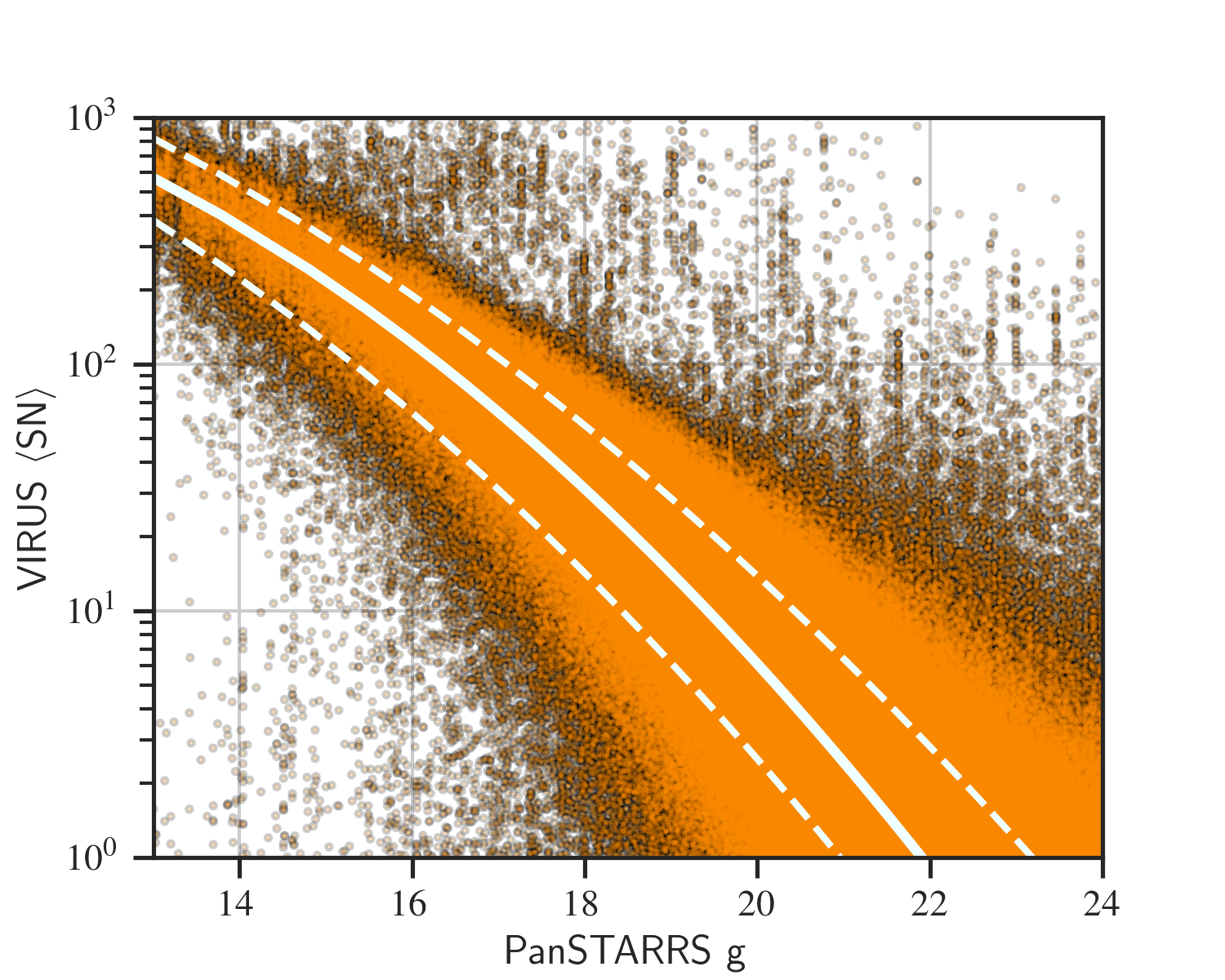}
	\caption{The relation between the PS1 DR2 $g$ magnitude (AB) versus the VIRUS S/N per 2~\AA\ pixel for the \numspec\ individual HETVIPS spectra.  The average relationship between $g$-magnitude and S/N is indicated by the solid white curve, with the 16$^{\rm th}$ and 84$^{\rm th}$ percentiles as dashed white curves.  At $g$=19.5, the average VIRUS S/N is roughly 10.}
	\label{fig:gmag_sn}
\end{figure}

We use the PS1 DR2 catalog (with a 5-$\sigma$ AB photometric depth in $grizy$ of 23.3, 23.2, 23.1, 22.3, and 21.3) as the basis for creating the HETVIPS object spectra.  For observations before 2021-June-1, we used the PS1 DR2 stacked catalog and for observations on/after 2021-June-1, we used the PS1 DR2 mean catalog.  We used the \texttt{astropy.mast} query system for both catalogs.  All the PS1 data used in this paper can be found in MAST: \dataset[https://doi.org/10.17909/s0zg-jx37]{https://doi.org/10.17909/s0zg-jx37}.  The two date ranges represent two discrete reduction efforts, and between the two efforts we could not recover our initial query mode and had to switch to the mean catalog.  We extract all PS1 DR2 catalog sources within the VIRUS field of view following \S\ref{sec:objextract}.  If less than 5\% of the light is covered by VIRUS fibers (assuming a point source model), the spectrum is excluded from further consideration.  We also require that the signal-to-noise ratio per 2~\AA\ pixel averaged over the PS1 $g$ filter is greater than one in the extracted VIRUS spectrum.  This last constraint may exclude emission-line objects that have weak continua. For the \numobs\ VIRUS exposures in HETVIPS, \numspec\ PS1 DR2 sources meet these criteria.  Because many primary targets have multiple visits, a given HETVIPS source can have multiple spectra. There are \numobj\ unique objects associated with the \numspec\ spectra.  For duplicate sources we retain the extracted spectrum with the highest S/N.  The relationship between PS1 $g$-magnitude and the VIRUS S/N is displayed in Figure \ref{fig:gmag_sn}.

\begin{figure}[t]
	 \includegraphics[width=1\columnwidth]{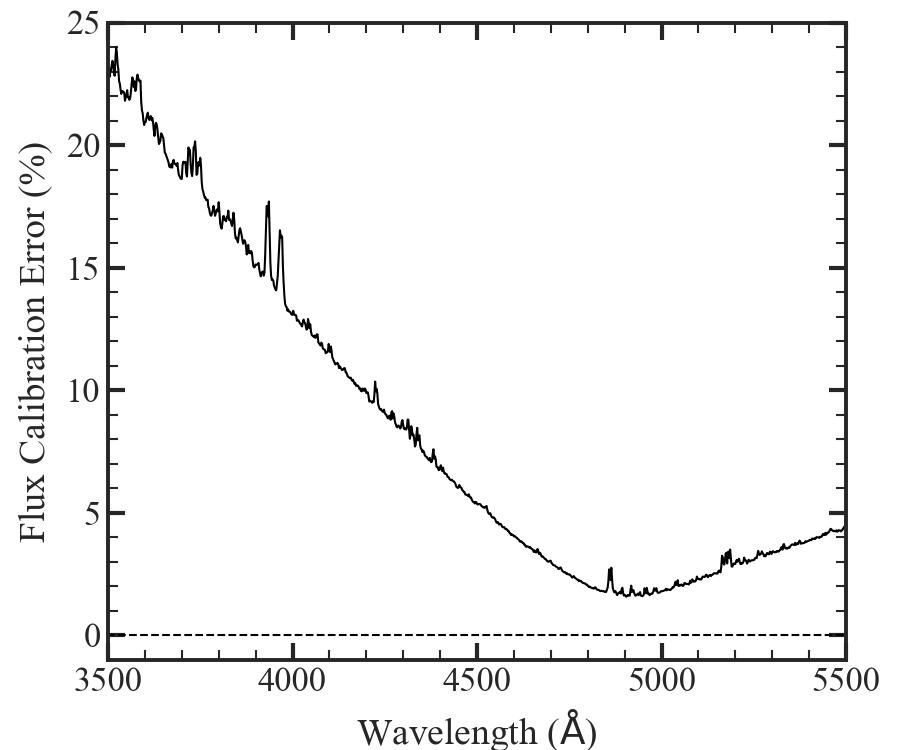}
	\caption{Repeat fields are used to compare the spectra of the same sources observed multiple times to quantify our flux calibration as a function of wavelength.  We use 39,526 repeat spectra and compare each spectrum to the highest signal-to-noise ratio source.  We then calculated the median absolute fractional difference using the highest signal-to-noise ratio source as the denominator.  Converting the median absolute fractional difference to a standard deviation includes the multiplication factor of $\approx$1.4826, and since we are searching for the flux calibration error of a single spectrum and not the differences of spectra we divide by $\sqrt{2}$.  This procedure produces our estimate of the flux calibration error as a function of wavelength.}
	\label{fig:flux_calibration_repeat}
\end{figure}

In the creation of the catalog, we perform an additional flux calibration step where a synthetic PanSTARRS $g$-magnitude from the VIRUS spectrum is normalized to the measured aperture $g$-magnitude from the PS1 DR2 catalog (recall that the aperture magnitudes from PS1 are corrected to a total flux using a PSF model).  This procedure ensures relatively accurate absolute flux calibration despite the sparse fiber coverage of the sources.  For resolved sources, a normalization to the PS1 aperture $g$-magnitude (PSF-corrected) may not be the most accurate approach and a more appropriate normalization would be to the PS1 Kron magnitude. However, any extrapolation of a resolved source's spectral information beyond the extracted 3\arcsec\ aperture used in Remedy should be done with care.   

\begin{figure}[t]
	 \includegraphics[width=1\columnwidth]{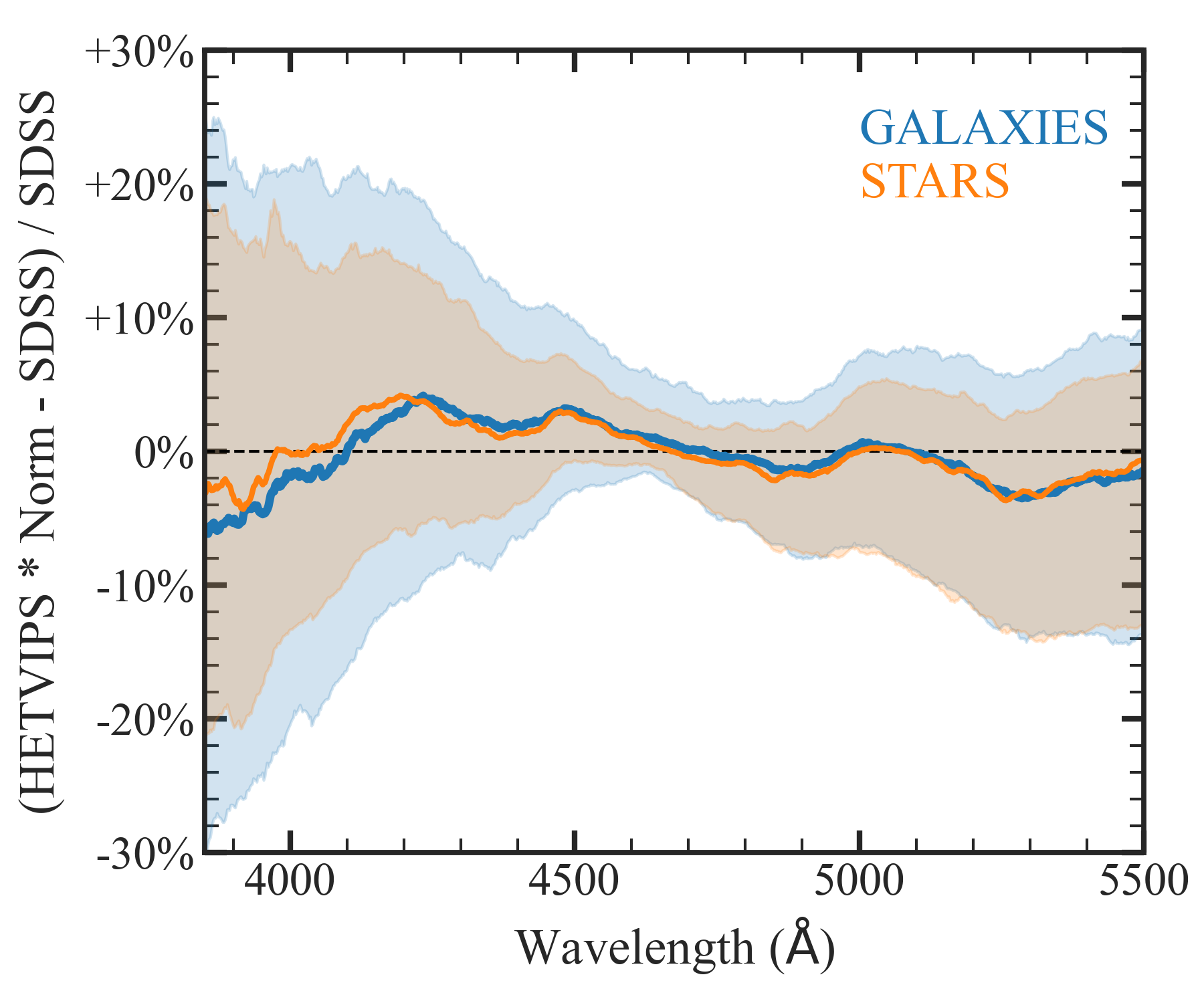}
	\caption{A comparison of the flux calibration of \numsdssgal\ galaxies and \numsdssstar\ stars found in both the SDSS and the HETVIPS surveys (limiting the HETVIPS catalog to only objects with VIRUS S/N $>$ 10).  After normalizing the HETVIPS spectra to \sdss\ measurements the median trend shows $<$5\% offset for both stars and galaxies.  The shaded regions cover the 16$^{\rm th}$-84$^{\rm th}$ percentile offsets for stars and galaxies (orange and blue, respectively), and represent the quadratic sum of errors for \sdss\ and HETVIPS flux calibration.}
	\label{fig:flux_calibration}
\end{figure}

Since many of our fields have repeat observations, we can internally calibrate our relative flux calibration.  We identified 12,989 sources in our unique object catalog that had a S/N $>$ 100, and matched them to our larger catalog of spectra including all of the repeat observations.  After excluding self matches, we constructed an array of 39,526 repeat spectra requiring that each have a S/N $>$ 20.  We calculated the median absolute fractional difference using the highest signal to noise ratio source as the denominator to estimate our flux calibration error as a function of wavelength;  Figure~\ref{fig:flux_calibration_repeat} shows the results. We find the lowest calibration errors occur near the pivot wavelength of the PS1 $g$ filter of around 2\% and rise to 4\% at 5500\AA\ and nearly 25\% at 3500\AA.

The HETVIPS catalog was matched to the Sloan Digital Sky Survey (\sdss) Data Release 16 (SDSS DR16) spectroscopic catalog \citep{sdssdr16} to measure the quality of our flux calibration externally.   HETVIPS covers roughly 26 deg$^2$ on sky and roughly one half of that falls within the footprint of SDSS DR16.  There were a total of \numsdssspec\ matches between the two catalogs.  Restricting the analysis to sources with S/N $>$ 10, there are \numsdssgal\ galaxies and \numsdssstar\ stars in common based on their \sdss\ classification.  Dividing the sample into those two categories reveals a good agreement between the two spectroscopic surveys after allowing a single normalization (Figure \ref{fig:flux_calibration}).

\section{Diagnose: Spectral Classification} \label{sec:diagnose}

To aid in the scientific use of this catalog, we developed a classification code, Diagnose \footnote{https://github.com/grzeimann/Diagnose}, which is described in the next subsection.  The code assigns one of four classifications for each source in the catalog (star, galaxy, quasar, or unknown), a redshift estimate for the galaxies and quasars, and a radial velocity estimate for the stars.  Similar to \cite{bolton2012}, the code is a spectral classification and redshift-finding algorithm via a ${\chi}^2$ minimization for linear combinations of principal component templates.

Diagnose uses a principal component analysis (PCA) to classify sources as stars, galaxies, quasars, or unknown. We used the templates of redrock \footnote{https://github.com/desihub/redrock-templates}, which are the same templates used for \sdss-IV \citep{Ross2020}.  The templates include 10 components for galaxies and four components for quasars.  Stars are further classified by type, with six components for B, A, F, G, K, M, and white dwarfs.  

\begin{figure}
	 \includegraphics[width=1\columnwidth]{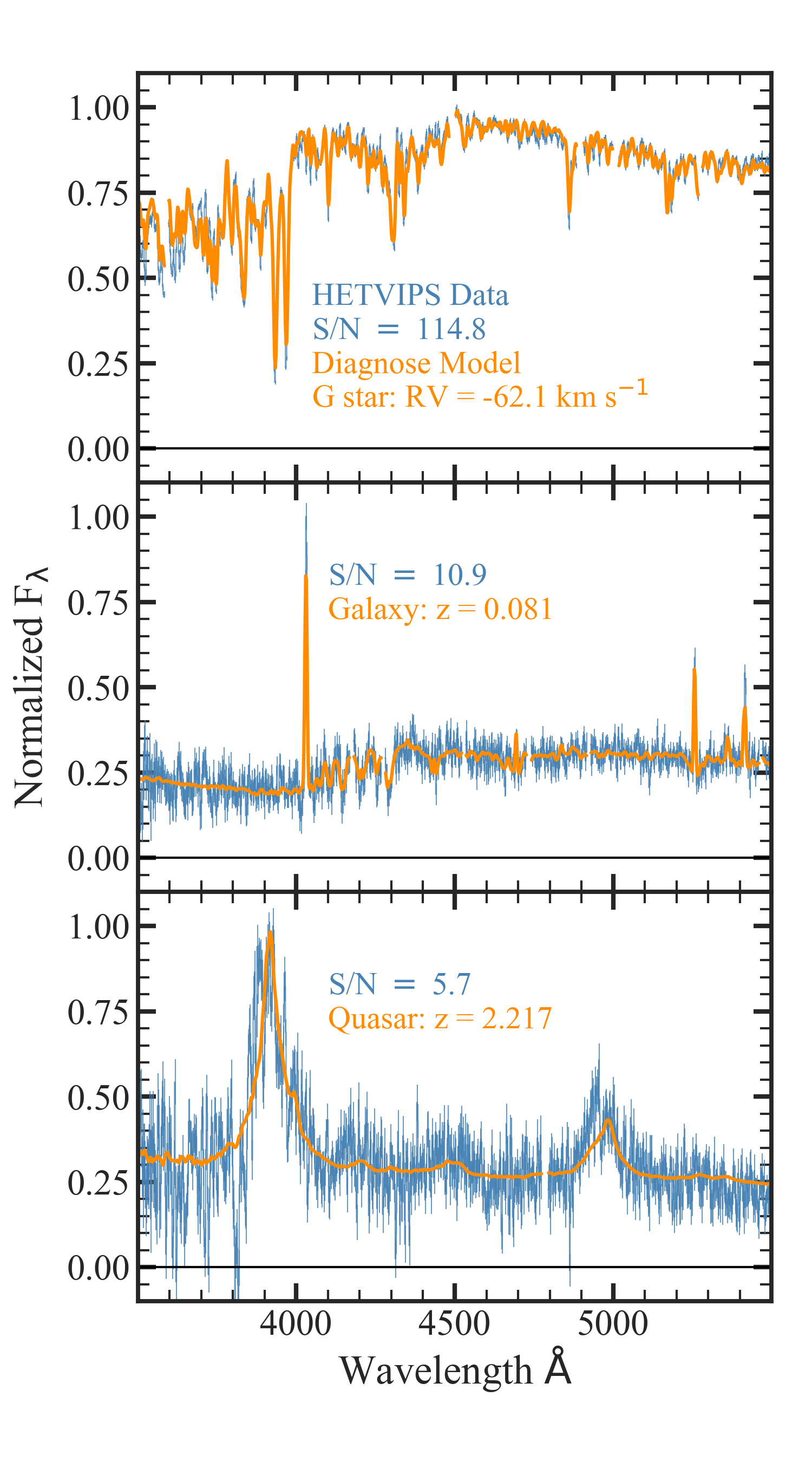}
	\caption{Three example spectra from the HETVIPS continuum catalog.  The VIRUS spectra are plotted as blue error bars and the best-fit model is shown in orange.}  The top spectrum is a G star, the middle spectrum is a low-redshift galaxy, and the bottom spectrum is a quasar.  The gaps in the spectra (e.g., on the short-wavelength wing of the [O~II] emisison line in the galaxy spectrum) are areas of masking where the fiber coverage of the source dropped below the threshold.
	\label{fig:example_spectra}
\end{figure}

We fit the stars over a range of radial velocities and the galaxies/quasars over a range of redshifts.  Before fitting, we construct a grid for each template across the range of redshifts/velocities shown in Table ~\ref{tab:fit}.  At each grid point we convolve the high-resolution templates from redrock to the spectral resolution of VIRUS.  For a given source spectrum, the reduced $\chi^2$ (\redchi) is used to evaluate the goodness of fit for each template and grid point.  The \redchi\ is just the original $\chi^2$ divided by the number of degrees of freedom (DoF), where DoF is equal to the number of unmasked data points in a spectrum minus the number of principle components for a template.  For each template, we refine the fit around the minimum \redchi\ velocity/redshift using the two neighboring grid points and fitting a second-order polynomial to the minimum.  For each spectrum, the classification code provides the best fit stellar type and velocity, the best fit galaxy and redshift, and the best fit quasar and redshift, as well as the best \redchi\ for each class.  The absolute minimum \redchi\ determines the best fit among the each of the template types.  We calculate the difference between the template with the best \redchi\ and the template with the second best \redchi; if this difference in values is larger than a statistical threshold ($\sqrt{2 / DoF}$), we classify the source as the best fit template (i.e., star, galaxy, or quasar).  If the difference is not larger than the threshold, we classify the source as unknown.  Figure \ref{fig:example_spectra} displays three example spectra from the HETVIPS continuum catalog, along with their best fit Diagnose model.

\begin{deluxetable}{r|c|c}
\label{tab:fit}
\tablecaption{Diagnose Velocity/Redshift Fit Parameters}
\tablehead{\colhead{Classification} & \colhead{Redshift/Velocity Range} & \colhead{Redshift/Velocity Step Size}}
\startdata
    Star & [-500, 500] km s$^{-1}$ & 50 km s$^{-1}$ \\
    Galaxy & [0.005, 0.470] & 0.001 \\
    Quasar & [0.10, 4.00] & 0.01 \\
    \enddata
\end{deluxetable}

\begin{figure}[h]
	 \includegraphics[width=1\columnwidth]{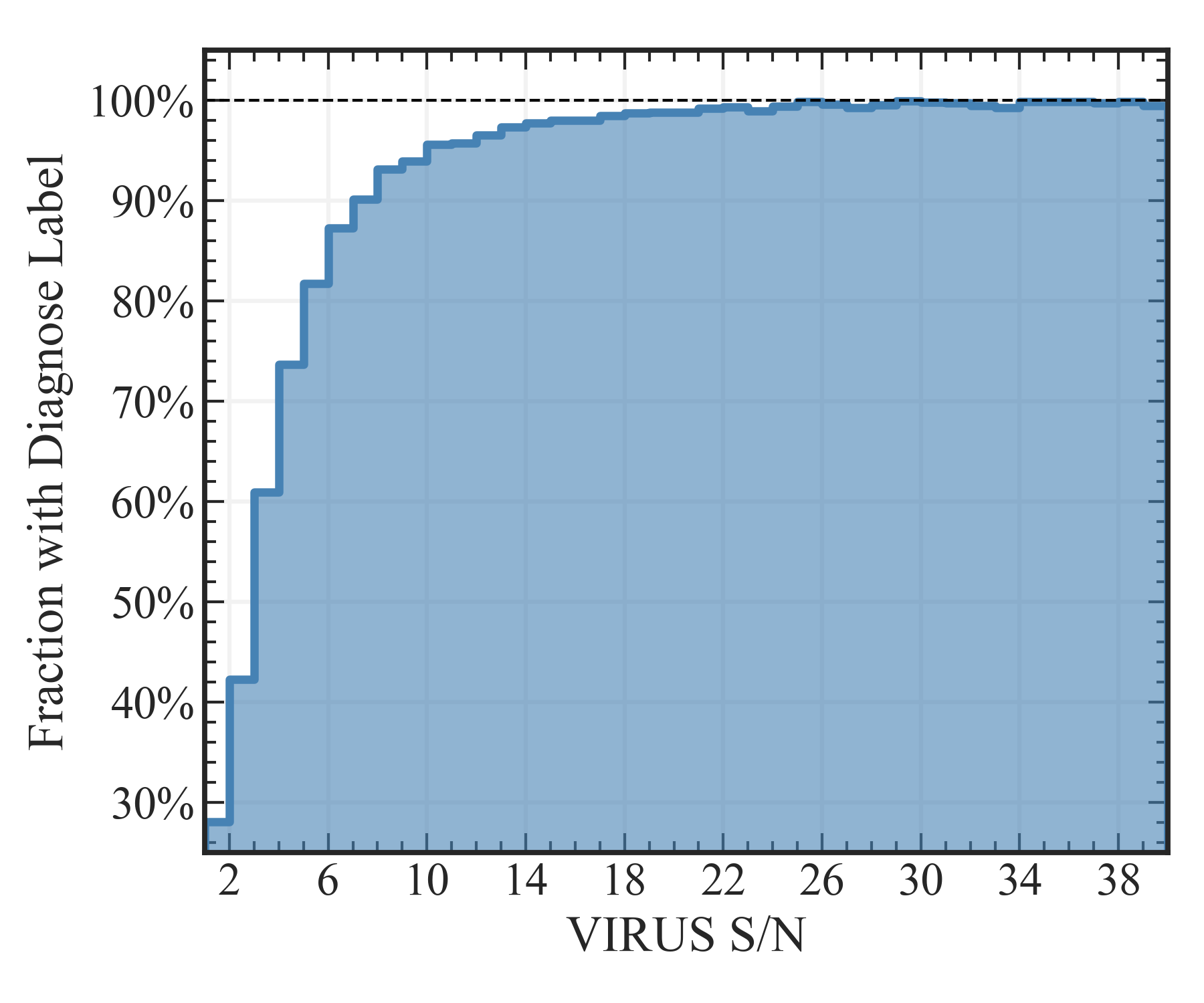}
	\caption{The fraction of sources with a classification from Diagnose versus VIRUS S/N.  A classification from Diagnose requires a the difference in ${\chi}^2$ values between the best fit label and the second best fit label to be above a given threshold.  At lower S/N, the ${\chi}^2$ values for multiple labels are too close to confidently distinguish between a galaxy, star, or quasar.}
	\label{fig:labeling}
\end{figure}

\begin{figure}
	 \includegraphics[width=1\columnwidth]{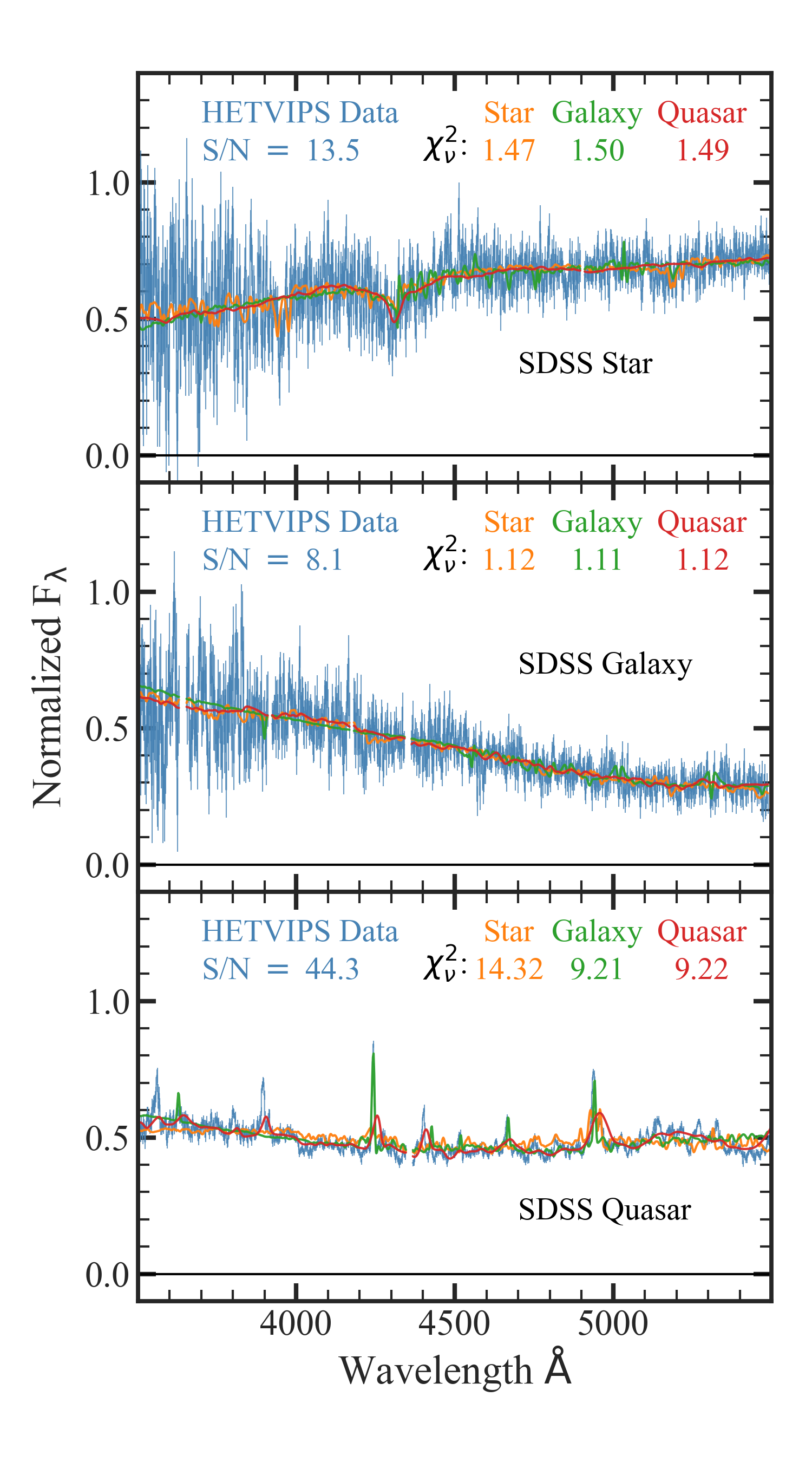}
	\caption{Three examples of spectra of moderate to high S/N that were classified as unknown.   The VIRUS spectra are plotted as blue error bars and the best-fit star, galaxy and quasar models are shown in orange, green and red, respectively. The SDSS classifications are star (top), low-redshift galaxy (middle), and low-redshift quasar (bottom).  Because the $\Delta$\redchi\ between the best fit VIRUS spectrum from each object is too small to meet the threshold, each of these sources is classified as unknown.}
	\label{fig:example_spectra_unknown}
\end{figure}

Our ability to classify a source is, as expected, strongly correlated with the S/N in the VIRUS spectrum.  Figure \ref{fig:labeling} presents the fraction of sources with a robust classification versus S/N. We also have notable exceptions in which the target has a moderate or high S/N, but the classification is unknown due to a small $\Delta$\redchi\ between the three labels.  Figure~\ref{fig:example_spectra_unknown} shows three examples of unknown spectra and the three best fit models from each group.

Also, just because a source has a classification does not necessarily means the label is correct. To determine the accuracy and efficacy of Diagnose, we compared the Diagnose classifications and redshifts to those from \sdss\ for the \numsdssspec\ sources that were spectroscopically observed in both surveys. 

\begin{figure}[t]
	 \includegraphics[width=1\columnwidth]{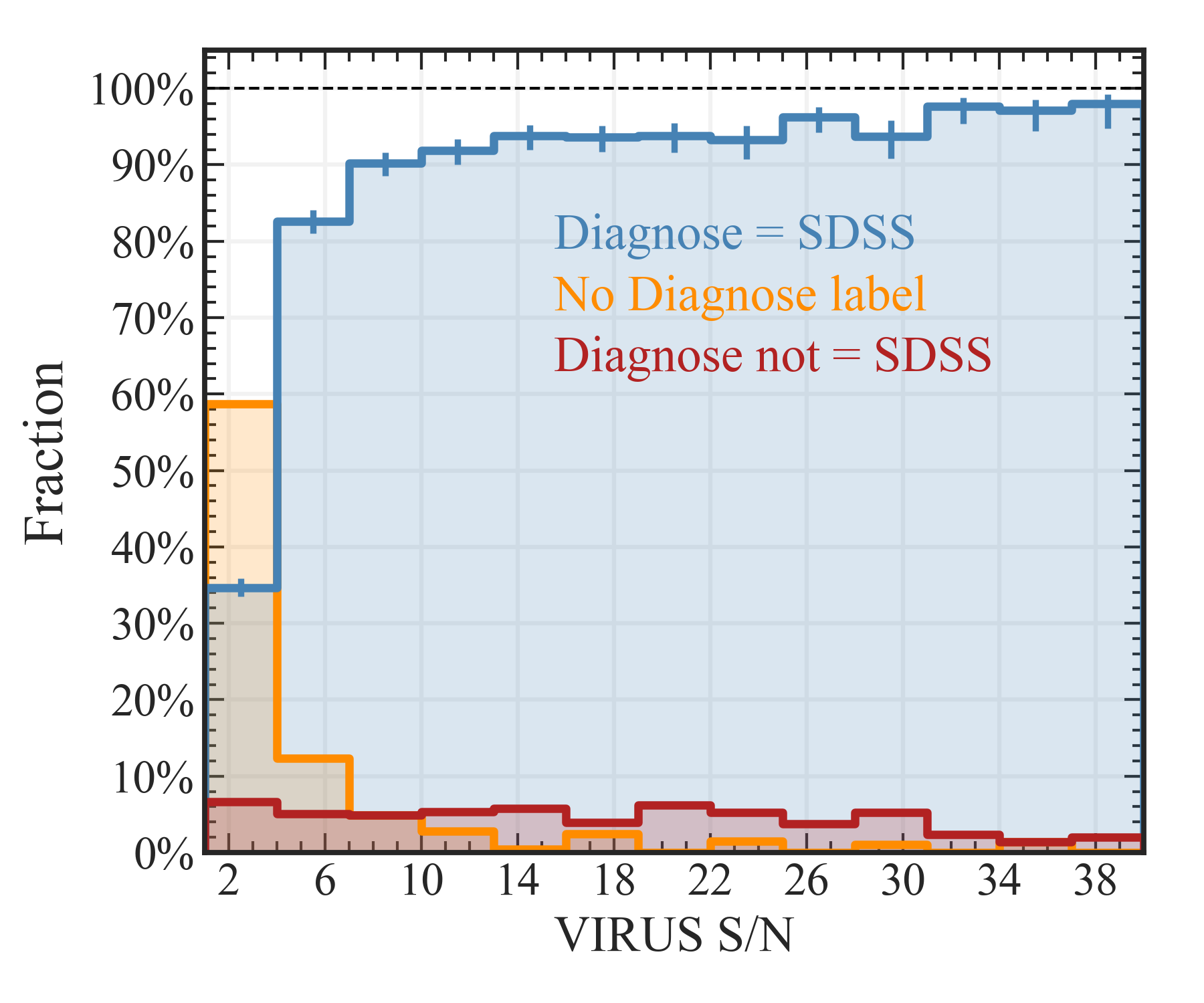}
	\caption{A comparison of the classifications from Diagnose to that of SDSS for stars, galaxies, and quasars as a function of VIRUS S/N per 2~\AA\ pixel.  The blue histogram with binomial error bars represents the fraction of sources with matching classifications, while the red histogram shows sources with mismatching classifications.  The orange histogram are the sources without a classification from Diagnose in which the the difference in ${\chi}^2$ values between the best fit label and the second best fit label was below the significance threshold.  The mismatches are not correlated with S/N but are rather uniformly distributed.}
	\label{fig:label_sdss}
\end{figure}

For the \numsdssspec\ sources spectroscopically observed in \sdss\ and HETVIPS, the labels assigned by the two surveys are in good agreement.  If we adopt the \sdss\ classification as truth, the Diagnose classification matched \sdss's classification for 96.9\%, 94.7\%, and 92.3\% for stars, galaxies, and quasars, respectively, when both SDSS and HETVIPS classifications were available.  Figure \ref{fig:label_sdss} shows the Diagnose classification versus SDSS classification as a function of VIRUS S/N.  The unknown sources are clustered at low S/N while the mis-classified sources are more evenly distributed across S/N.  

\begin{figure}[t]
	 \includegraphics[width=1\columnwidth]{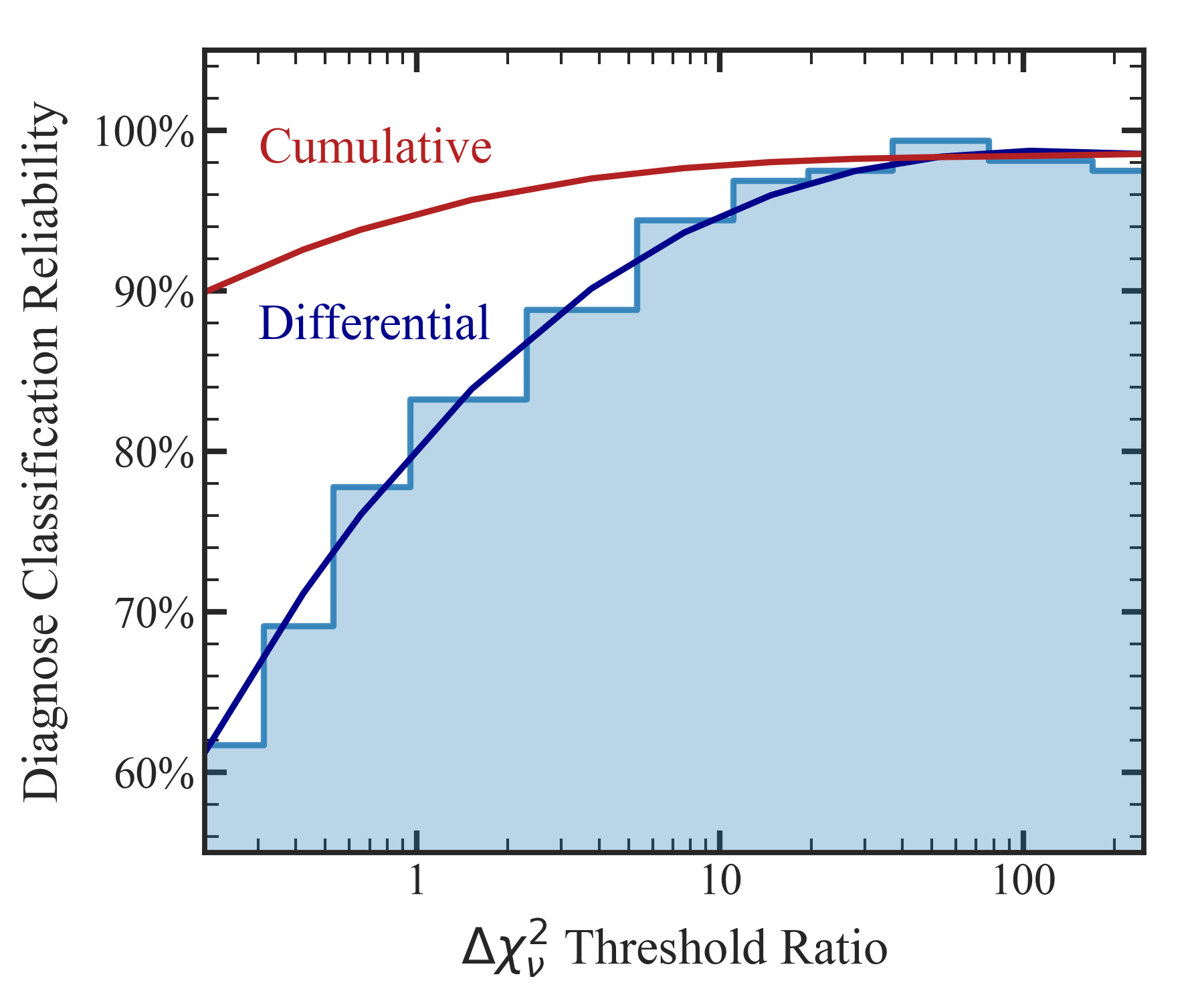}
	\caption{A relation between the classification reliability from Diagnose (i.e., classification matched SDSS) as a function of difference of the best fit model \redchi\ minus the second best fit model \redchi\ divided by the statistical threshold ($\sqrt{2 / DoF}$). The blue histogram represents the actual label reliability, while the blue curve is a 3$^{\rm rd}$ order polynomial fit to the blue histogram.  The cumulative reliability is indicated in red, which would be the expected reliability of the total catalog for a given threshold ratio cut.  At 0.2 times the current \redchi\ threshold, the entire sample is expected to be 90\% reliable in its labels, while at the current threshold value the reliability is around 95\%.  The ``differential'' reliability in blue drops rapidly below 1 as a function of threshold choice.}
	\label{fig:class_thresh}
\end{figure}

We also investigate the reliability of our classification as a function of the difference of the best fit model \redchi\ minus the \redchi\ of the second best fit model, divided by the statistical threshold ($\sqrt{2 / DoF}$).  Figure \ref{fig:class_thresh} presents the Diagnose classification reliability (i.e., the classification matches SDSS) versus the $\Delta$\redchi\ threshold ratio.  The reliability of our classification at exactly a threshold ratio of one is roughly 80\% and the reliability of a catalog with a threshold ratio greater than one is about 95\%.

The most common mis-classification occurs when Diagnose labels the source as a galaxy while \sdss\ identifies the source as a quasar.  The limited wavelength coverage of VIRUS compared to \sdss\ means that single, narrow lines can easily be reproduced by both a galaxy template basis and a quasar template basis. Since the galaxy basis has 10 components rather than the 4 of the quasar, the extra degrees of freedom (even when accounted for in the \redchi) result in a Diagnose mis-classification.  This case of mis-classification produces an incorrect redshift assessment as the upper boundary of the galaxy redshift range is $z$=0.47, and Ly$\alpha$ is often confused with [O~II].

\begin{figure*}[t]
	 \includegraphics[width=2\columnwidth]{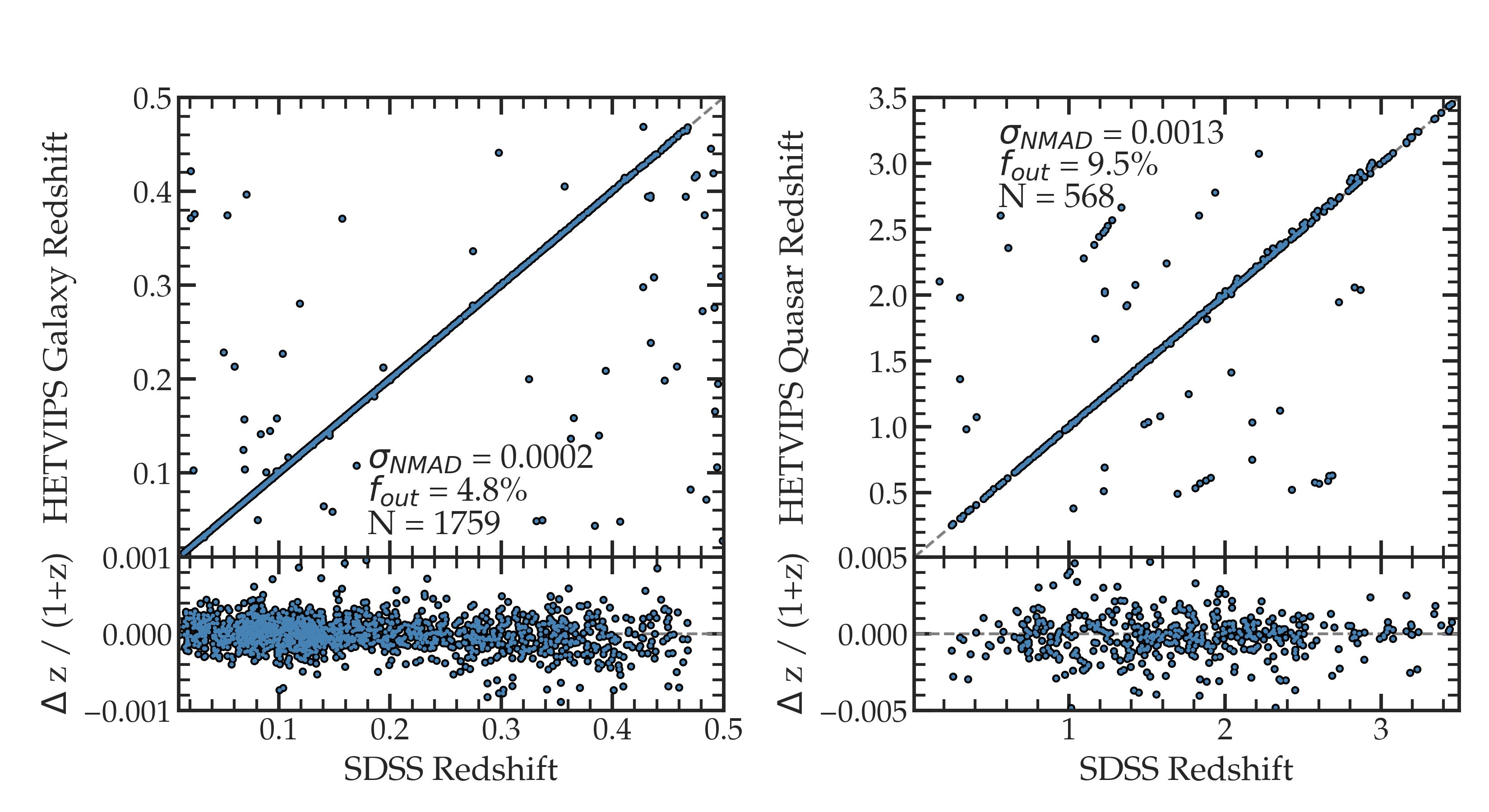}
	\caption{The galaxy and quasar redshifts from Diagnose are compared to those from SDSS. The normalized median absolute deviation (NMAD) and outlier fraction are calculated as in Momcheva et at. (2016).  The redshifts for the galaxies are highly reliable; the quasar measurements, while generally accurate, do suffer from a number of catastrophic failures.}
	\label{fig:sdss_z}
\end{figure*}

We compare the redshifts derived from Diagnose and SDSS in Figure \ref{fig:sdss_z}.  The galaxy redshifts are in excellent agreement with only 5\% outliers, and a redshift standard deviation of ${\Delta}z = 0.0002$.  The quasar redshifts in Figure \ref{fig:sdss_z} are also in good agreement, but include nearly 10\% outliers and a larger standard deviation of ${\Delta}z = 0.0013$.  

\subsection{Catalog Breakdown}
\label{sec:catbreakdown}

After processing the \numobj\ unique HETVIPS continuum sources through  Diagnose, \numstars\ objects are classified as stars, \numgal\ as  galaxies, \numqso\ as quasars, and \numunknown\ as unknown. Diagnose also  produces the best fit stellar type.  The program classified \numbstar\ objects as B stars, \numastar\ as A stars, \numfstar\ as F stars,  \numgstar\ as G stars, \numkstar\ as K stars, \nummstar\ as M stars,  \numwstar\ as white dwarfs.  We caution interpretation of the spectral type distribution, as there are known issues with the subgroup  classifications from Diagnose (see \S\ref{sec:knownissues}).

\subsection{Known Issues}
\label{sec:knownissues}

As stated in \S\ref{sec:diagnose},  $\sim$~97\% of sources classified as stars by Diagnose were also classified as stars by SDSS.  Diagnose also produces the best fit stellar type: B, A, F, G, K, M, or W (white dwarf).  We inspected each population of stars within the subgroups and found the labels to be somewhat unreliable.  Flux calibration is discussed in \S\ref{sec:objextract}, and with individual errors as large as 20-25\% in the blue, it is quite easy for stars to move from one subgroup label to another due to mis-calibration, limited wavelength coverage, and PCA flexibility.  Also, even though one subgroup may be listed as the best fit, the second best \redchi\ may not be substantially different than the best fit subgroup label.  

We plan to improve Diagnose's ability to assign stellar types in the future; in this paper, the label of a HETVIPS star is likely to be correct but, the sub classification should be independently verified by the user.

Since sources are extracted individually with no consideration for contribution from their neighboring systems, we expect that some of our object spectra include small contributions from their neighbors.  Also, we perform a local sky subtraction, which can be effected by bright stars and extended galaxies.  Sky subtraction and neighbor contribution effects are strongest in crowded stellar fields.  Over subtraction of the ``sky'' can lead to negative fluxes, incorrect continua shape, and mis-classifications (especially for stellar subgroups).  

Although \remedy\ employs a fiber masking effort to identify fiber spectra effected by bad columns, cosmic rays, or amplifier issues, with \nfibers\ fiber spectra in HETVIPS DR1 there are a number of potential issues that may be unidentified.  Objects extracted from fibers with unidentified issues typically exhibit odd continua and exaggerated absorption features. The subgroup of F stars seem to contain most of these issues.  This behavior is not an intrinsic feature of F stars, but the F star PCA basis has the greatest flexibility to model the object when it is truly a star and suffers from these data processing issues.  In future releases, we plan to more thoroughly capture odd behaviors in object spectra and provide a comprehensive flag system for quality assurance.

\section{Data Products}
\label{sec:datadist}

The overview of the HETVIPS data products is provided by an ASCII table, which has one line per source.  The
table contains celestial coordinates, PS1 photometry, VIRUS S/N, \redchi\ for each classification, the redshift for each classification, and the best fit redshift (see Table \ref{table:ascii}).

\begin{table*}[t]
 
\centering
\begin{tabular}{||>{\columncolor[gray]{0.8}}c l l||} 
 \hline
 Column Name & Description & Units  \\ [0.5ex] 
 \hline\hline
 objID & HETVIPS Object ID & string  \\
 RA & PanSTARRS1 Right Ascension (J2000)  & degree \\
 Dec & PanSTARRS1 Right Ascension (J2000)  & degree \\
 shotid & VIRUS Shot ID \{DATE\}\_\{OBSNUM\}   & string \\
 gmag & PanSTARRS1 aperture $g$ magnitude (PSF corrected to total flux) & AB magnitude \\
 rmag & PanSTARRS1 aperture $r$ magnitude (PSF corrected to total flux) & AB magnitude \\
 imag & PanSTARRS1 aperture $i$ magnitude (PSF corrected to total flux) & AB magnitude \\
 zmag & PanSTARRS1 aperture $z$ magnitude (PSF corrected to total flux) & AB magnitude \\
 ymag & PanSTARRS1 aperture $y$ magnitude (PSF corrected to total flux) & AB magnitude \\
 sn & VIRUS S/N averaged over PS1 $g$ filter & float \\
 barycor & Barycentric velocity correction & m s$^{-1}$ \\
 mjd & Modified Julian Date for the middle of the VIRUS exposure & day \\
 exptime & Exposure time for the VIRUS observation & s \\
 chi2\_star & \redchi\ for best fit stellar PCA template (and velocity) & float \\
 chi2\_galaxy & \redchi\ for best fit galaxy PCA template (and redshift) & float \\
 chi2\_qso & \redchi\ for best fit quasar PCA template (and redshift) & float \\
 z\_star & Best fit stellar template redshift  & float \\
 z\_galaxy & Best fit galaxy template redshift  & float \\
 z\_qso & Best fit quasar template redshift  & float \\
 z\_best & Best fit template redshift (if classification is unknown, z\_best = -999) & float \\
 classification & STAR, GALAXY, QSO, or UNKNOWN & string \\
 stellartype & Best fit templates for the stellar PCA (B, A, F, G, K, M, C, or W) & string\\[1ex]
 \hline
\end{tabular}
\caption{Description of the ASCII catalog columns.  Stellar type W is white dwarf.}
\label{table:ascii}
\end{table*}

Diagnose produces a FITS file with several extensions containing information about each source. These extensions include: the best fit models (star, galaxy, and quasar), the classification, \redchi\ (for stars, galaxies, and quasars), statistical threshold for the $\Delta$\redchi, redshifts (star, galaxy, and quasar), the best fit stellar type, as well as a table of information include PS1 photometry and sky position. The catalog is broken into twenty FITS files to keep the individual size to roughly 500Mb. 

We describe in greater detail each of the extensions in numerical order:
\begin{enumerate}
\item The Diagnose models and VIRUS spectra.  It is a 3-dimensional array that is N$_{obj}$ long, six rows down, and 1,036 columns deep.  The 1,036 columns are the length of the spectra starting at 3470\AA\ in intervals of 2\AA\ ending at 5540\AA.  The six rows are the best fit Diagnose star model, galaxy model, quasar model, VIRUS spectrum, VIRUS error spectrum, and VIRUS fiber weight array (see \S\ref{sec:objextract}), respectively.  The header of the primary extension describes this information, as well including the wavelength information.
\item The Diagnose classifications.  We use numerical representation for the classifications; specifically, stars are 1, galaxies are 2, quasars are 3, and unknown sources are 4.  All extensions are N$_{obj}$ long with the same ordering as extension 1.
\item The \redchi\ values for the best fit models of the three groups: stars, galaxies, and quasars, respectively.
\item The \redchi\ threshold for each source in the FITS file.  The \redchi\ statistical threshold is $\sqrt{2 / DoF}$.  This threshold was used to compare the best fit Diagnose \redchi\ model to the second best fit, and if the difference was bigger than the threshold, a unique classification was determined.
\item The best fit redshifts/velocities for each source.  The first column is the best fit velocity in km/s, while the second and third column are the best fit redshift for the galaxy and quasar groups, respectively.
\item The best fit stellar type: B, A, F, G, K, M, or W (white dwarf).  
\item A binary table about each source in the FITS file.  The columns of the table include the HETVIPS object ID, right ascension (J2000), declination (J2000), VIRUS shot ID, the PS1 photometry, the average VIRUS S/N in the spectrum, the barycentric velocity correction (not already applied to the spectra), the date, and the VIRUS shot's exposure time.
\end{enumerate}

We do not provide the database of \nfibers\ VIRUS fiber spectra from which the catalog was generated.  That product remains internal to the HET community.

\subsection{Data Distribution}

The ASCII catalog and the FITS files are available publicly at \url{https://web.corral.tacc.utexas.edu/hetdex/HETVIPS/}.  We suggest downloading the ascii catalog to investigate the sky positions of the \numobj\ sources in the FITS files.  If you want to investigate the spectra, the easiest method is to download the FITS files and loop through each file to execute your desired program. 

\section{Science Cases}
\label{sec:science}

The HETVIPS spectroscopic dataset is valuable for a range of scientific programs.  We highlight just a few science cases for consideration.  The HETVIPS spectra, covering 3470 \AA\ to 5540 \AA, can be used to determine a star's effective temperature, gravity, and metallicity (e.g., [Fe/H]) as demonstrated by \citet{Hawkins2020a}.  Future studies could look for hyper velocity stars, extremely metal poor stars, or investigate stellar metallicity of the Milky Way along many sight-lines through the galaxy.  The Balmer lines (e.g., H$\beta$) can be used to determine the mass accretion rate of a young stellar objects (YSOs) to help understand how stars form and how their circumstellar disks evolve with time as was already done with a small subset of VIRUS parallel observations in Willett et al. (in prep).  The HETVIPS spectra can also provide a way to classify a number of supernova objects found by photometric surveys as shown in \citet{Vinko2023}.  Finally, \citet{Liu2023} illustrated that VIRUS spectra can also be used to investigate the spatially resolved pre-explosion local environments of known supernovae.

\section{Summary } 
\label{sec:summary}

We present HETVIPS DR1 derived from VIRUS parallel data taken on the HET from 01 Jan 2019 to 31 Mar 2023.  The raw database consists of approximately \nfibers\ spectra obtained in \numobs\ observations and although most fibers are dominated by sky, we construct and publicly distribute the HETVIPS DR1 Continuum Catalog, which consists of \numobj\ unique object spectra located over a wide region on the celestial sphere.  The spectra are classified using a code called \texttt{Diagnose}, with each object divided into one of four groups (star, galaxy, quasar, or unknown).  The information contained in the HETVIPS DR1 Catalog, including spectral classifications and redshifts, can be used to address a wide variety of scientific projects ranging from the chemical history of the Galaxy to the local environment of extragalactic supernovae.

\begin{acknowledgments}
    
We thank the staff at the Hobby Eberly Telescope  for making this project possible.

The authors acknowledge the Texas Advanced Computing Center (TACC) at The University of Texas at Austin for providing computing resources that have contributed to the research results reported within this paper. URL: http://www.tacc.utexas.edu

The Pan-STARRS1 Surveys (PS1) and the PS1 public science archive have been made possible through contributions by the Institute for Astronomy, the University of Hawaii, the Pan-STARRS Project Office, the Max-Planck Society and its participating institutes, the Max Planck Institute for Astronomy, Heidelberg and the Max Planck Institute for Extraterrestrial Physics, Garching, The Johns Hopkins University, Durham University, the University of Edinburgh, the Queen's University Belfast, the Harvard-Smithsonian Center for Astrophysics, the Las Cumbres Observatory Global Telescope Network Incorporated, the National Central University of Taiwan, the Space Telescope Science Institute, the National Aeronautics and Space Administration under Grant No. NNX08AR22G issued through the Planetary Science Division of the NASA Science Mission Directorate, the National Science Foundation Grant No. AST–1238877, the University of Maryland, Eotvos Lorand University (ELTE), the Los Alamos National Laboratory, and the Gordon and Betty Moore Foundation.

Some of the data presented in this paper were obtained from the Mikulski Archive for Space Telescopes (MAST) at the Space Telescope Science Institute. The specific observations analyzed can be accessed via \dataset[https://doi.org/10.17909/s0zg-jx37]{https://doi.org/10.17909/s0zg-jx37}. STScI is operated by the Association of Universities for Research in Astronomy, Inc., under NASA contract NAS5–26555. Support to MAST for these data is provided by the NASA Office of Space Science via grant NAG5–7584 and by other grants and contracts.

Funding for SDSS-III has been provided by the Alfred P. Sloan Foundation, the Participating Institutions, the National Science Foundation, and the U.S. Department of Energy Office of Science. The SDSS-III web site is http://www.sdss3.org/.

SDSS-III is managed by the Astrophysical Research Consortium for the Participating Institutions of the SDSS-III Collaboration including the University of Arizona, the Brazilian Participation Group, Brookhaven National Laboratory, Carnegie Mellon University, University of Florida, the French Participation Group, the German Participation Group, Harvard University, the Instituto de Astrofisica de Canarias, the Michigan State/Notre Dame/JINA Participation Group, Johns Hopkins University, Lawrence Berkeley National Laboratory, Max Planck Institute for Astrophysics, Max Planck Institute for Extraterrestrial Physics, New Mexico State University, New York University, Ohio State University, Pennsylvania State University, University of Portsmouth, Princeton University, the Spanish Participation Group, University of Tokyo, University of Utah, Vanderbilt University, University of Virginia, University of Washington, and Yale University.

\end{acknowledgments}

\vspace{5mm}
\facilities{HET, MAST (Pan-STARRS)}

\software{Astropy \citep{2013A&A...558A..33A, astropy2018, astropy2022},  
          scipy \citep{2019arXiv190710121V},
          numpy \citep{harris2020array}}

\bibliography{bibliography}{}
\bibliographystyle{aasjournal}

\end{document}